\pdfoutput=1
\documentclass{JINST}

\title{A Radiation-Hard Dual Channel 4-bit Pipeline for a 12-bit 40 MS/s  ADC Prototype with extended Dynamic Range for the ATLAS Liquid Argon Calorimeter Readout Electronics Upgrade at the CERN LHC}

\author{J. Kuppambatti$^a$, J. Ban$^b$, T. Andeen$^b$, P. Kinget$^{a,1}$, G. Brooijmans$^{b,2}$\\
\llap{$^a$}Columbia University, Dept. of Electrical Engineering\\
  New York, New York, USA\\
\llap{$^b$}Columbia University, Nevis Laboratories\\
  Irvington, New York, USA\\
  E-mail: \llap{$^1$}\email{kinget@ee.columbia.edu} ,  \llap{$^2$}\email{gusbroo@nevis.columbia.edu}}

\abstract{The design of a radiation-hard dual-channel 12-bit 40 MS/s pipeline ADC with extended dynamic range is presented, for use in the readout electronics upgrade for the ATLAS Liquid Argon Calorimeters at the CERN Large Hadron Collider. The design consists of two pipeline A/D channels with four Multiplying Digital-to-Analog Converters with nominal 12-bit resolution each. The design, fabricated in the IBM 130 nm CMOS process, shows a performance of 68 dB SNDR at 18 MHz for a single channel at 40 MS/s while consuming 55 mW/channel from a 2.5 V supply, and exhibits no performance degradation after irradiation. Various gain selection algorithms to achieve the extended dynamic range are implemented and tested.}


\begin{document}

\section{Introduction} \label{sec:Intro}
This article describes the design and performance of a radiation-hard dual channel 40 MS/s pipeline ADC prototype with extended dynamic range. The future evolution of this ADC is intended for the upgraded electronics in the ATLAS Liquid Argon Calorimeter readout. The current prototype is used to establish the analog performance of the pipeline, to study the radiation tolerance of the ADC design and to determine the optimal gain selection procedure 
to be implemented in the future version of the chip.

The Large Hadron Collider (LHC) at CERN in Geneva has been operational for physics research since 2010 \cite{lhc}. In this proton-proton collider, designed to  operate at  center of mass energies of 7--14 TeV, the high luminosities (> $10^{34}~\mathrm{cm}^{-2}\mathrm{s}^{-1}$)
produce an intense radiation environment that the detectors and their electronics must withstand.

The ATLAS detector  \cite{jinst} is a multi-purpose apparatus constructed to explore the new particle physics regime opened by the LHC. The energy of the created electrons and photons is measured by a sampling calorimeter technique that uses liquid argon as its active medium. The front-end electronic readout of the
ATLAS liquid argon (LAr) calorimeter consists of a combined analog and digital processing system \cite{FE}. To record the large dynamic range signals from the liquid argon calorimeter  with high precision and to limit noise, a substantial portion of the electronic system is located on the ATLAS detector itself.

The LAr calorimeters of the ATLAS experiment have functioned with excellent reliability since installation
in 2006 \cite{operations}. In 2012, they  played a pivotal role in the observation of the Higgs boson, particularly in the diphoton and four
electron channels \cite{higgs}. Looking toward the future operation of the LAr calorimeters, there are several constraints: the existing front-end electronics limit the granularity, bandwidth
and latency in the on-line event selection (level-1 trigger). Without modifications, the peak instantaneous luminosity of $3\times10^{34}~\mathrm{cm}^{-2}\mathrm{ s}^{-1}$, expected in the next few years, would force substantial increases in trigger thresholds. Additionally, the on-detector electronics, which are exposed to substantial radiation, are complex and contain many technologies, resulting in many potential opportunities for failure. The front-end electronics were qualified for radiation levels corresponding to 10 years of LHC operations \cite{existingRad}. The high luminosity running of the LHC (HL-LHC) \cite{pahse1loi}, with instantaneous luminosities of $5\times10^{34}~\mathrm{cm}^{-2}\mathrm{s}^{-1}$ and an integrated luminosity of $3000~\mathrm{fb}^{-1}$, will exceed these design qualifications.

\begin{figure}[h]
\includegraphics[width=35pc]{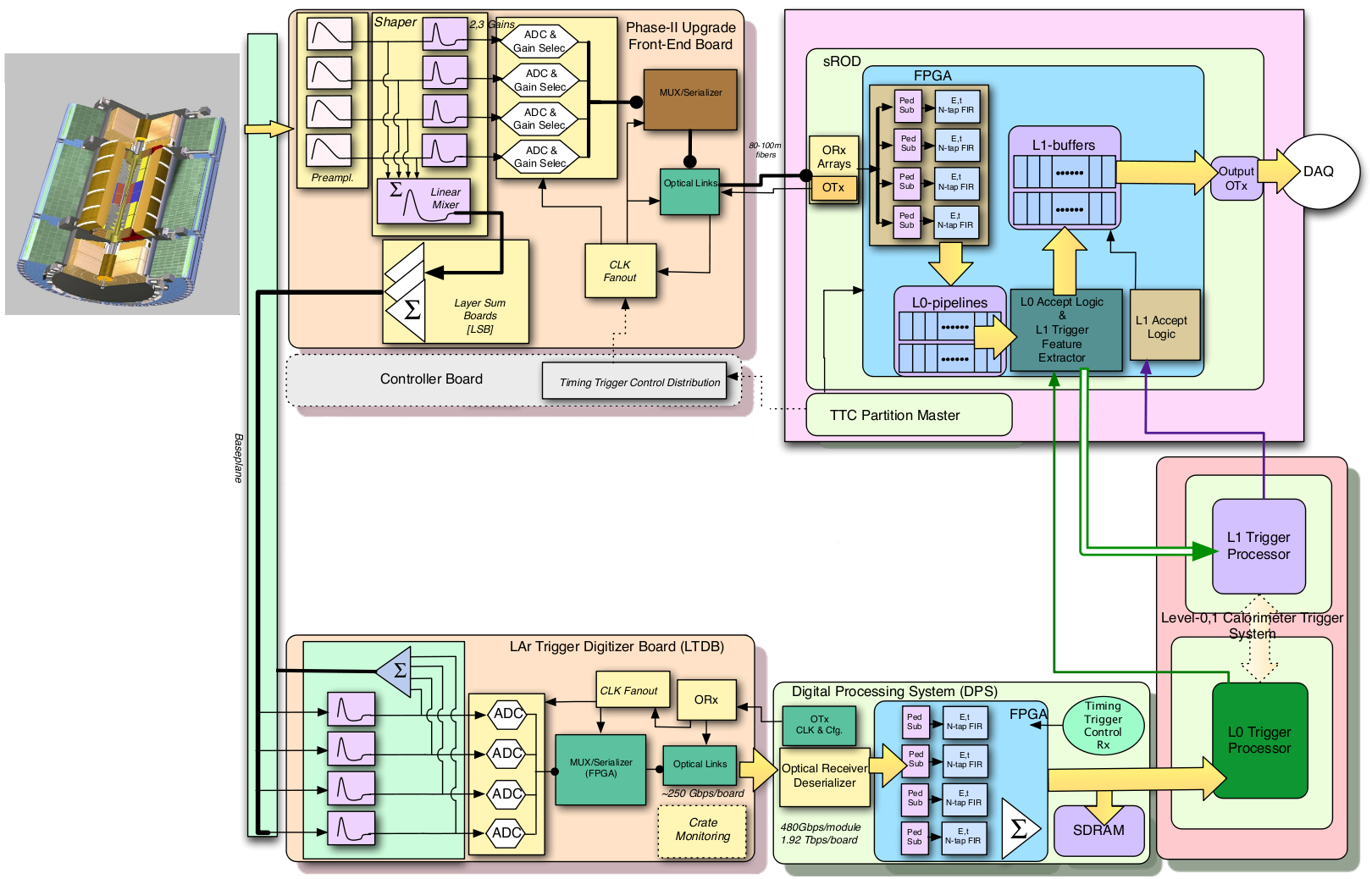}
\caption{\label{phase2block}Block diagram for the proposed ATLAS Phase-II electronics upgrade. The ADC appears in the upper left box (FEB) and the lower left box (LTDB). }
\end{figure}

For the HL-LHC, planned for 2022, it will be necessary to upgrade all of the 1524 front end boards (FEBs), preparing the detector for the expected $3000~\mathrm{fb}^{-1}$. This replacement allows the design of a more flexible system that utilizes the full precision and granularity of the LAr calorimeters at trigger level and removes bandwidth and latency constraints. The simpler, ``free running'' architecture chosen results in an effectively infinite pipeline and bandwidth with little or no latency. This can be seen schematically in Figure \ref{phase2block}, which shows the path of the signal from the detector to the data acquisition (DAQ) system.  The new FEBs  (upper left) will digitize the analog signals for all LHC bunch crossings. The data will be sent to an off-detector digital system, the Read Out Driver (ROD, upper right), that will provide input for a new, fully digital level-1 trigger system. A proposed level-0 trigger may then repurpose the Liquid Argon Trigger Digitizer Boards (LTDB, lower left) and off-detector Digital Processing System (DPS, lower right) from their prior use in the level-1 trigger. Technically, the new FEBs require 40 MS/s digitization for all channels over a 16-bit dynamic range within the existing power, cooling and space constraints, and in a radiation environment.  Thus, an important component of this upgrade is the creation of a suitable analog-to-digital converter (ADC). The ADC is shown in Figure \ref{phase2block} on the FEB,  receiving the signal from the preamplifier-shaper, and also on the LTDB.

\section{System Architecture}

\subsection{System Specifications} \label{sec:sys_specs}
\begin{figure}
\begin{center}
		\includegraphics[scale=0.45]{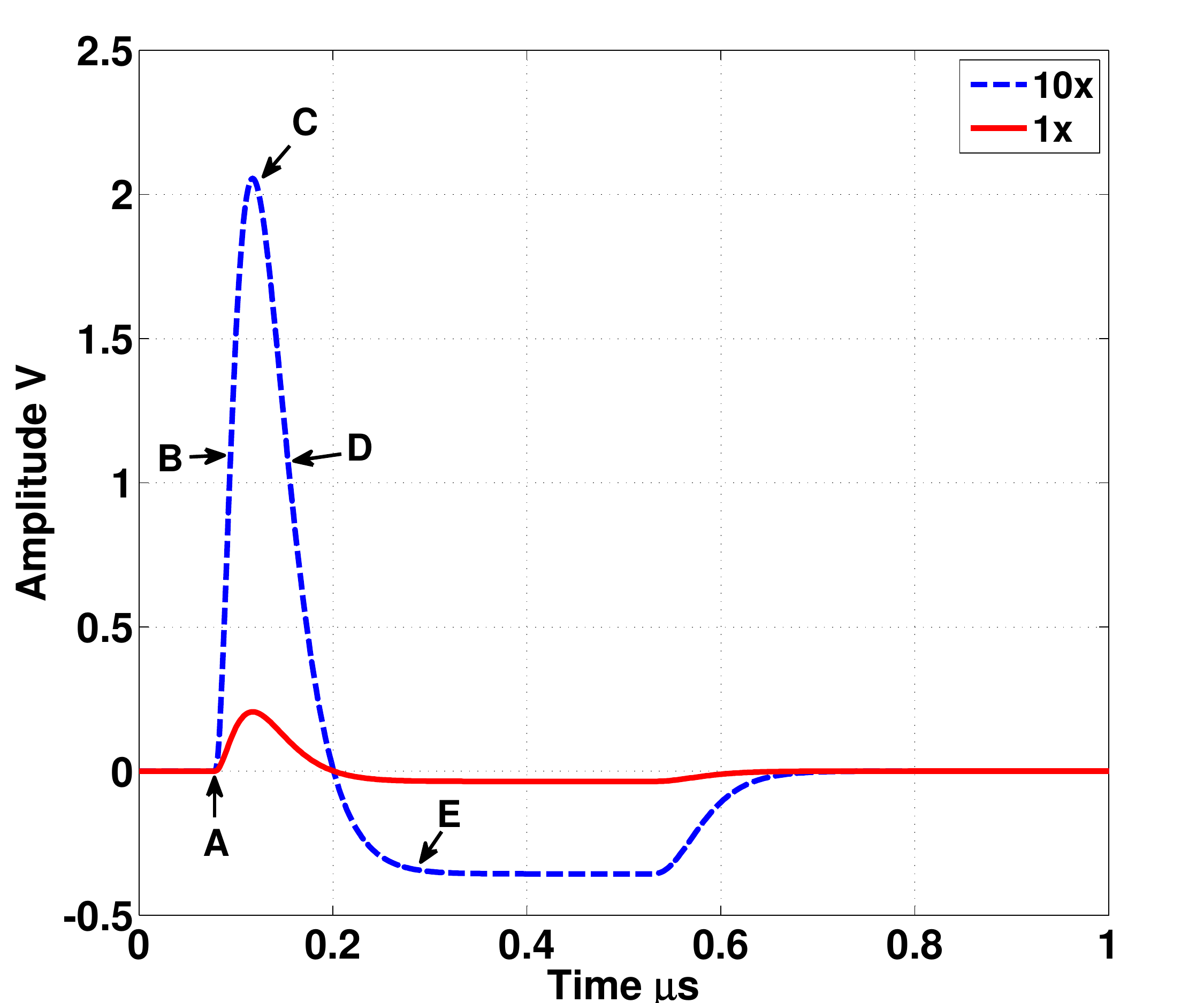}
		\caption{Pulse Shape with 1x (solid line) and 10x gain (dashed line).}
		\label{fig:pulse_shape}
\end{center}
\end{figure}

The ADC is designed to digitize the analog signal from a single calorimeter cell. The signal, a triangular pulse about 500~ns long, is first sent to a preamplifier and then split and again amplified by shaper chips to produce three pulses at overlapping linear gain scales, with gain ratios of approximately ten. Each signal is subject to a fast bipolar $CR-(RC)^{2}$ shaping function with $\tau$~=~RC~=~20~ns, resulting in a pulse shape similar to that shown in Figure \ref{fig:pulse_shape} for the 1x and 10x gain channels. This pulse is sampled five times for digitization: typically one sample before the signal starts (point $A$ in Figure \ref{fig:pulse_shape}), one during the rising edge ($B$), one at the peak ($C$), one during the falling edge ($D$) and one during the undershoot phase ($E$). With additional processing and calibration, the energy deposited in a particular calorimeter cell is determined based on these five samples.

The LAr calorimeter electronics upgrade must meet the following specifications:
\begin{itemize}
\item The calorimeter signals must be sampled and digitized at (minimally) a frequency of 40 MHz.
\item The energy deposited in each calorimeter cell must be measured with a precision of better
than 0.25\% at high energy. 
\item A dynamic range of  approximately 16 bits is needed to cover the energy
range of interest, from a lower limit of approximately 50~MeV set by pile-up noise \cite{FE} up to a maximum of 3~TeV.
\item 128 ADC channels must fit on a board 490 mm $\times$ 409.5~mm.
\item The ADC must use less than 100 mW per channel.
\item The ADC must be radiation tolerant to $\sim$ 1 MRad and have low single event effect sensitivity.
\end{itemize}

This combination of requirements is not currently available commercially. The chip discussed in this paper is a proof-of-principle demonstration of a design meeting these specifications.

\subsection{Radiation Tolerance}
The radiation in the ATLAS detector is dominated by  secondary particles produced by interactions of the primary particles with the detector elements. As a result, at the electronics location, the energies are rather low (less than a few GeV), the fluxes are high, and the direction of the radiation fields is homogeneous. The dominant radiation flux consists of photons and neutrons. Also contributing to the radiation are charged hadrons (mainly protons and pions). A high level of reliability of the electronics must be maintained during the estimated ten years of operation of the experiment. As mentioned in Section \ref{sec:sys_specs}, the ADC needs to be radiation tolerant to $\sim$1 MRad.
The decrease in Total Ionizing Dose (TID) effects in thin oxide gates of MOS devices has been shown in \cite{rad_tol_ref1, rad_tol_ref2}. Although technology scaling has resulted in the gate oxide getting thinner, and hence less susceptible to TID damage, the Shallow Trench Isolation (STI) oxide of modern CMOS technologies ultimately limits the radiation tolerance of conventional CMOS circuits. It is possible with Hardness-By-Design layout techniques to eliminate this limitation and eventually push the radiation tolerance of circuits to the high level allowed by the thin gate oxide. The radiation hardness of IBM's 130~nm CMOS process for digital design has been shown \cite{rad_tol_ref3} for both thin oxide and thick oxide MOS devices to a few tens of MRads. A key decision while implementing the ADC prototype is the choice of the power supply. The thin-oxide 130~nm devices have a rated supply of 1.2~V. The ADC chip is required to interface with the  shaper chip, which will probably be implemented in a SiGe process with a supply of 3.3~V, and hence interfacing this signal to a 1.2~V domain will require attenuation and/or complicated level shifting. Signal attenuation from a 3.3~V regime to a 1.2~V regime will increase the noise requirements of the ADC. Also, implementing analog circuits in a 1.2~V supply is more challenging due to the reduced voltage headroom, which makes it infeasible to cascade MOS devices on top of each other. To address these issues, the prototype ADC is implemented on a supply of 2.5~V using thick-oxide devices. Although it was seen \cite{rad_tol_ref3} that thin oxide devices are more radiation tolerant, thick oxide devices are still comfortably tolerant at the radiation levels required for this application ($\sim$ 1 MRad). The allowable input signal swing of the ADC is 2.4~V$_{p-p}$ differential, which relaxes the noise requirements of the ADC. Measurement results in a radiation environment, on a previous design of a sample and hold amplifier in the same technology, showed that standard good analog layout practices were sufficient to achieve the necessary radiation tolerance \cite{Hastings_ref}. The measured radiation tolerance of this chip is discussed in Section \ref{sec:meas_results}.

\subsection{Prototype Implementation}
\label{sec:arch}
\begin{figure}
	\begin{center}
		\includegraphics[scale=0.375]{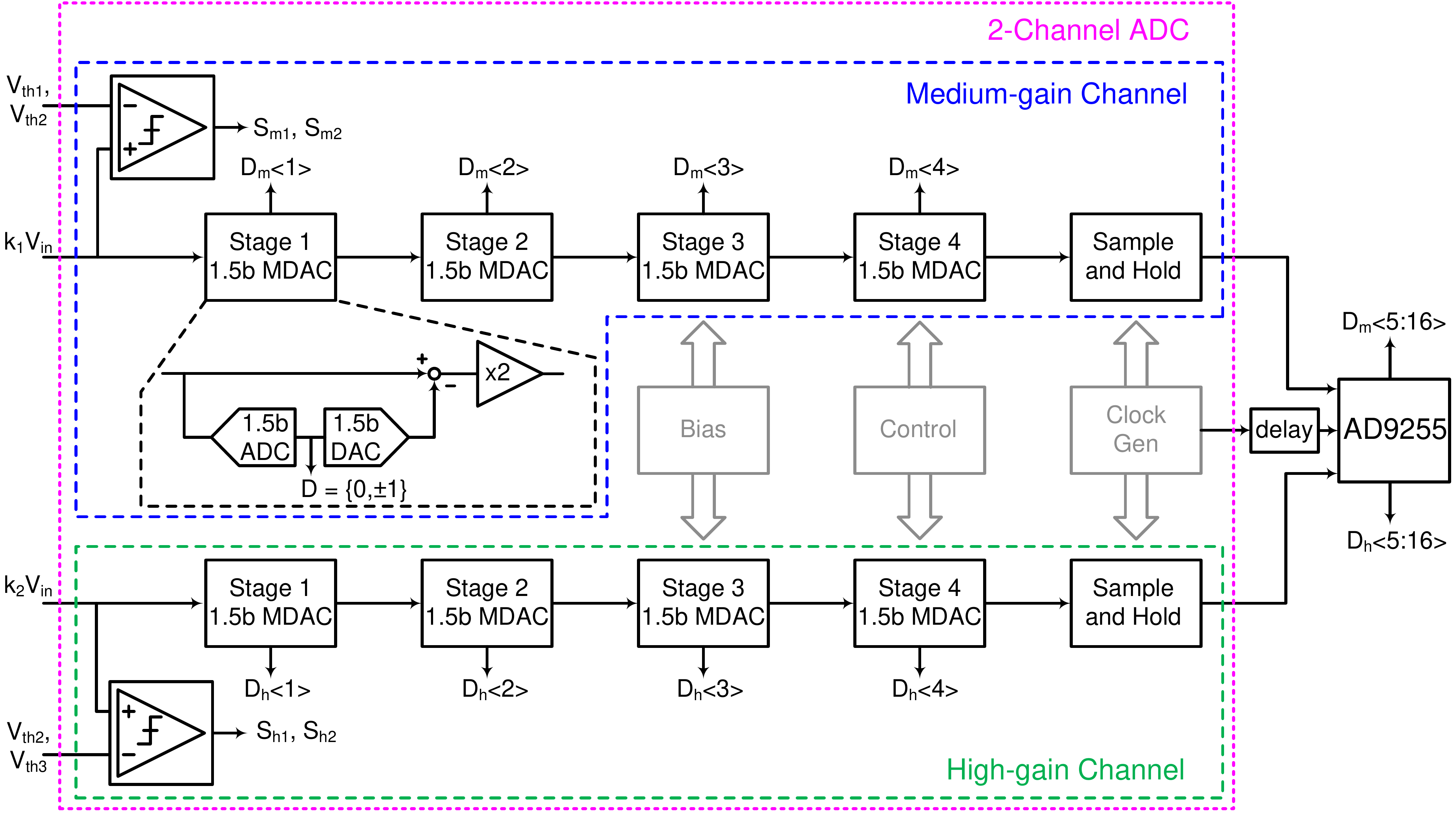}
		\caption{Prototype architecture.}
		\label{fig:chip_arch}
	\end{center}
\end{figure}

The goals of the current prototype are to establish the analog performance of the pipeline stages, to study the radiation tolerance of the ADC design and to determine the optimal gain selection procedure to be implemented in the future version of the chip. The system prototype architecture, shown in Figure \ref{fig:chip_arch}, consists of two identical 12-bit 40 MS/s Pipeline ADC channels. The two channels, namely Medium Gain and High Gain, are fed with $k_1V_{in}$ and $k_2V_{in}$ respectively, where the gain factors $k_1$ and $k_2$ are provided by off-chip drivers and are nominally one and ten respectively. This two-channel system prototype, with four MDAC (Multiplying Digital-to-Analog Converters) stages each, provides enough flexibility to test various gain selection algorithms to increase the effective system dynamic range.

\begin{figure}
	\begin{center}
		\includegraphics[scale=0.5]{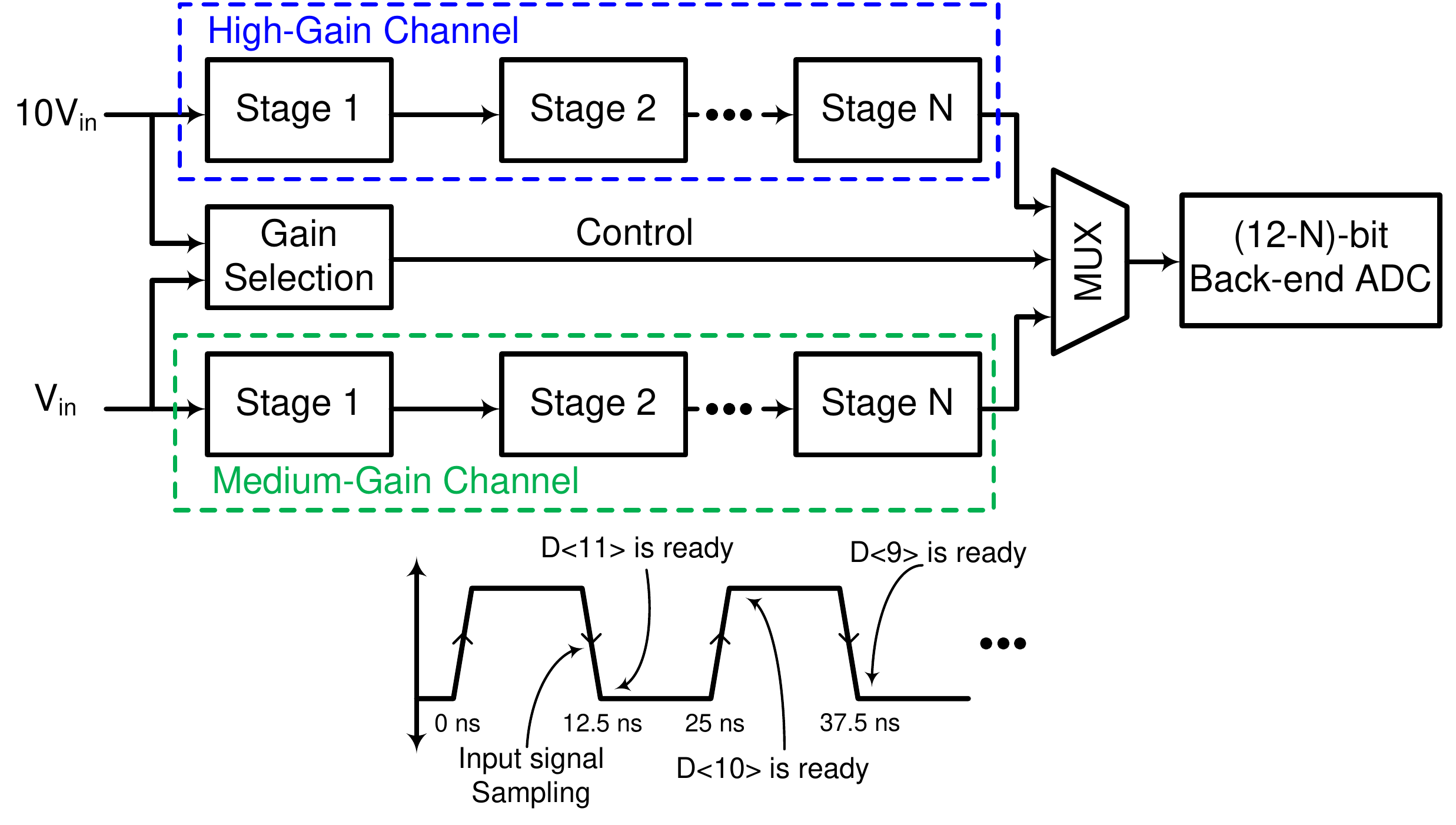}
		\caption{Gain selection system: D<11:0> is the reconstructed 12-bit ADC output; D<11> is the bit from Stage 1, D<10> from Stage 2 
                 etc. }
		\label{fig:GS_block}
	\end{center}
\end{figure}

Figure \ref{fig:GS_block} shows the general block diagram of a system based on the current ADC prototype that implements gain selection. For simplicity, a two-channel system is shown. The ADC prototype is a pipeline architecture so the final reconstructed digital output is available only {\it after} a delay of a few clock cycles. Gain selection requires the detection of saturation in the high gain channel and then switching to the lower gain channel. The high and low gain channels are separated by a factor of ten, similar to the existing architecture. 

Each ADC channel in the current prototype consists of four MDAC stages with a gain of two each. Each MDAC stage resolves one bit, followed by a times two residue amplification. The three possible output codes (``1.5 bits'') allow for the digital error correction. The analog residue of the fourth and final MDAC stage needs to be further resolved to 8-bit accuracy to determine the final 12-bit ADC word. This analog residue is fed to an on-chip Sample and Hold amplifier, which drives an external commercial 12-bit ADC (AD9255 \cite{AD9255_ref}) for further digitization, i.e.~to determine the eight least significant bits. In the current prototype, for simplicity, all MDACs are identical. Power scaling will be done in the final design to reduce the total chip power consumption. The implementation details of the ADC are further explained in the following sections.

\subsubsection{MDAC Implementation}
\begin{figure}
	\begin{center}
		\includegraphics[scale=0.55]{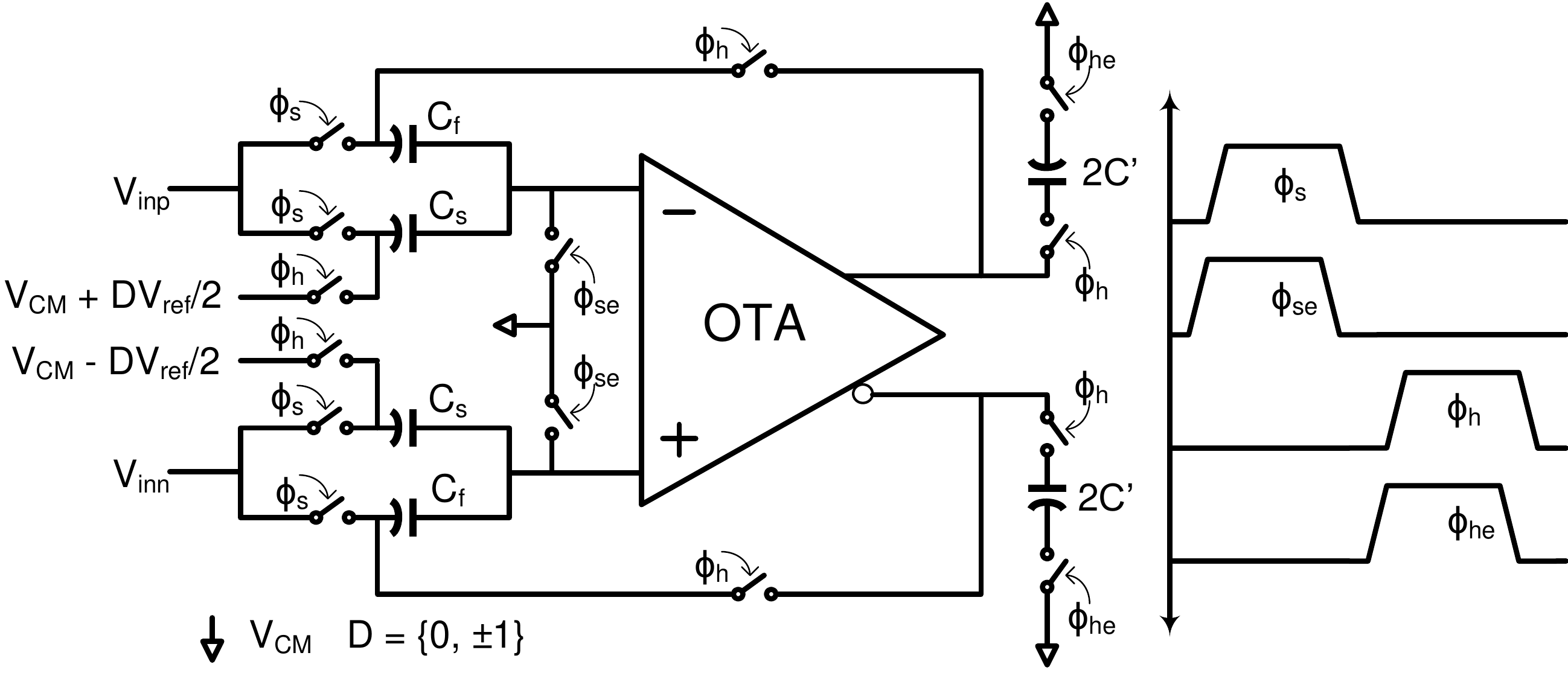}
		\caption{1.5b MDAC stage (subADC not shown); $\phi_s/\phi_h$ - sample/hold phase; $\phi_{se}/\phi_{he}$ - advanced version of $\phi_s/\phi_h$; $V_{CM}$ - common-mode voltage; $V_{ref}$ - reference voltage; $D$ - subADC decision.}
		\label{fig:MDAC}
	\end{center}
\end{figure}

Figure \ref{fig:MDAC} shows the architecture of the 1.5-bit MDAC stage used in the pipeline ADC \cite{stg_arch}.
The MDAC consists of a flip-around architecture with the input $V_{in}$ sampled onto the sampling capacitors $C_s$ and the flip-around capacitors $C_f$, during the sampling phase $\phi_s$. Nominally, $C_s$ and $C_f$ are equal. In order to improve the linearity of the input sampling network, bottom-plate sampling, which uses an advanced version $\phi_{se}$ of the sampling phase, is implemented. At the beginning of the hold phase $\phi_h$, depending on the subADC decision $D$, the reference $V_{ref}$ is appropriately connected to the capacitors, thus performing the required MDAC operation given by

\begin{equation}
        V_{out} = 2V_{in} - DV_{ref}
        \label{eqn:MDAC_opn}
\end{equation}
where $D = {0,\pm1}$ is the subADC decision.

\begin{figure}
	\begin{center}
		\includegraphics[scale=0.55]{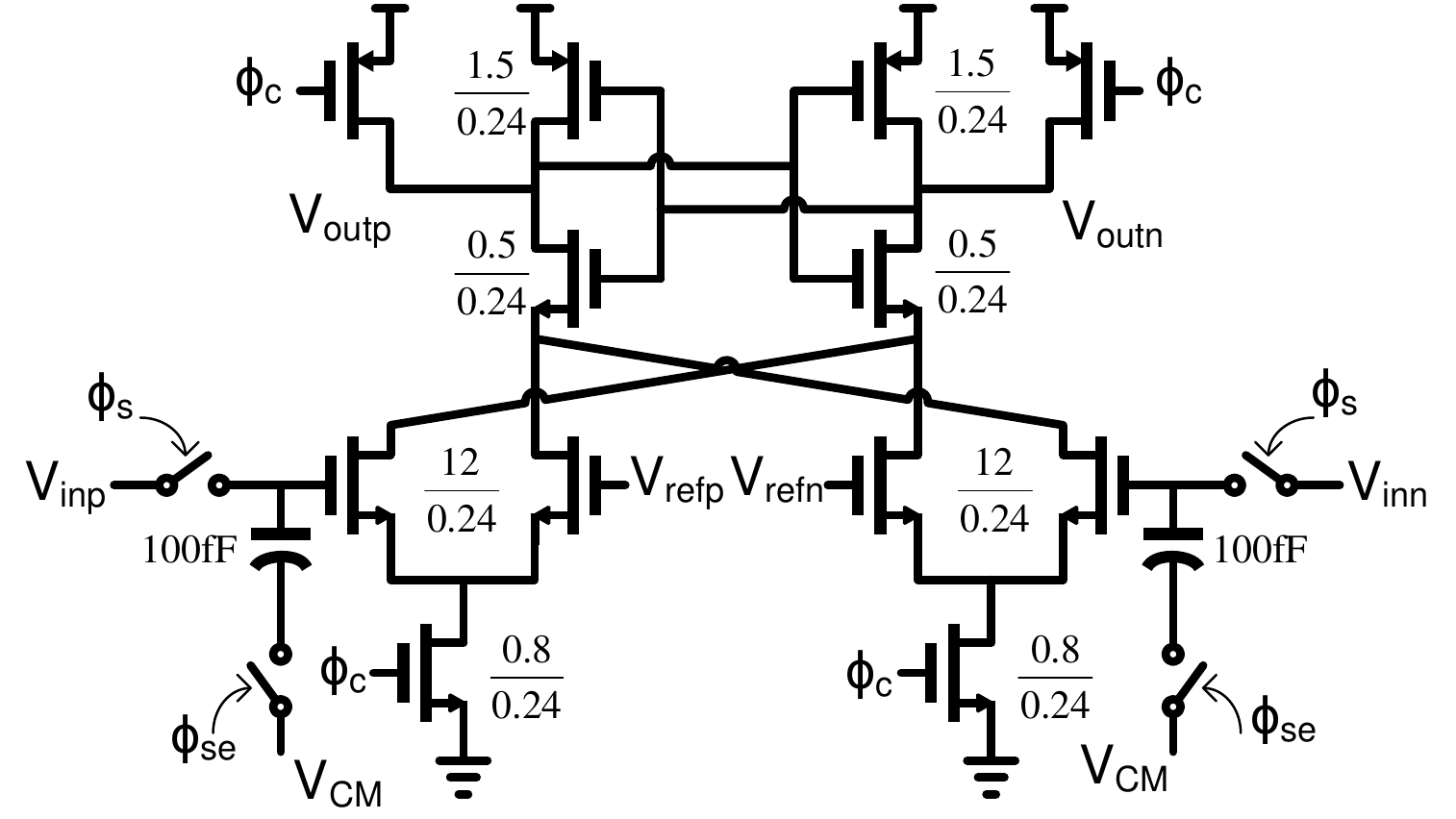}
		\caption{A single subADC unit; $\phi_s$ - sample phase; $\phi_\mathit{se}$ - advanced version of $\phi_s$; $\phi_\mathit{c}$ - subADC comparison phase; $V_\mathit{CM}$ - common-mode voltage; $V_\mathit{refp}/V_\mathit{refn}$ - reference voltage. All dimensions are in $\mu$m.}
		\label{fig:subADC}
	\end{center}
\end{figure}

Fig.~\ref{fig:subADC} shows a single unit of the subADC consisting of a flash comparator and an input sampling network. The separate sampling network of the subADC obviates the need for a Sample-and-Hold amplifier at the ADC input. The flash comparator is sized to comfortably meet the subADC offset requirements. Additional logic to perform gain calibration is included on-chip. MiM (Metal-insulator-Metal) capacitors are used because of their better matching and higher density in the given process. In order to avoid device reliability issues in a radiation environment, clock gate boosting is not employed for the input sampling switches, which also avoids the need for triple well devices. Instead, bottom-plate sampling, along with appropriately sized thick-oxide transmission-gate switches, was found to satisfy the 12-bit linearity requirements at 40 MS/s.

\subsubsection{OTA}
\begin{figure}
	\begin{center}
		\includegraphics[scale=0.5]{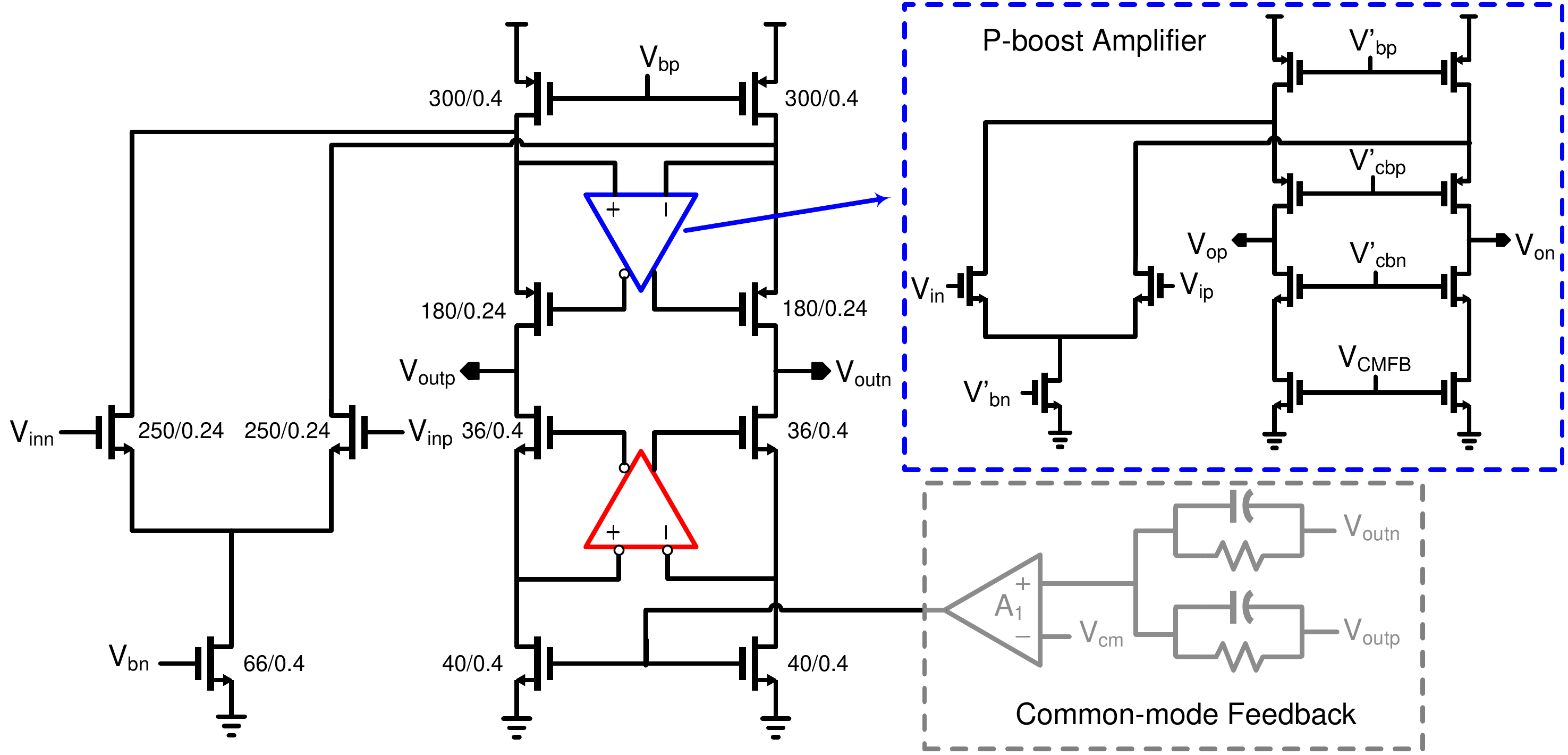}
		\caption{Folded-cascode OTA; $V_{cm}$ - common-mode voltage; $V_{CMFB}$ - common-mode feedback voltage; $V_{bn},V'_{bn}/V_{bp},V'_{bp}$ - NMOS/PMOS bias voltage; $V'_{cbn}/V'_{cbp}$ - NMOS/PMOS cascode bias voltage; All dimensions in $\mu$m.}
		\label{fig:OTA}
	\end{center}
\end{figure}
Figure \ref{fig:OTA} shows the fully differential OTA (Operational Transconductance Amplifier) used in the MDAC (bias circuits not shown), along with the transistor sizes. The OTA consists of a single stage gain-boosted folded cascode amplifier with an NMOS input pair. The load capacitance of the OTA forms the dominant pole. Care is taken to make sure the pole-zero doublet formed by the gain-boosting amplifiers falls beyond the unit-gain frequency (UGB) of the OTA \cite{OTA_ref}. The boost amplifiers are also implemented as single-stage folded cascode amplifiers. The common-mode feedback amplifier $A_1$ is implemented as a PMOS differential pair with diode-connected NMOS loads. The OTA is designed for a DC gain of 80~dB and a UGB of 450~MHz, thus providing enough margin for the targeted 12-bit performance at 40 MS/s.

\subsubsection{Digital Gain Calibration} \label{sec:dig_calib}
\begin{figure}
	\begin{center}
		\includegraphics[scale=0.8]{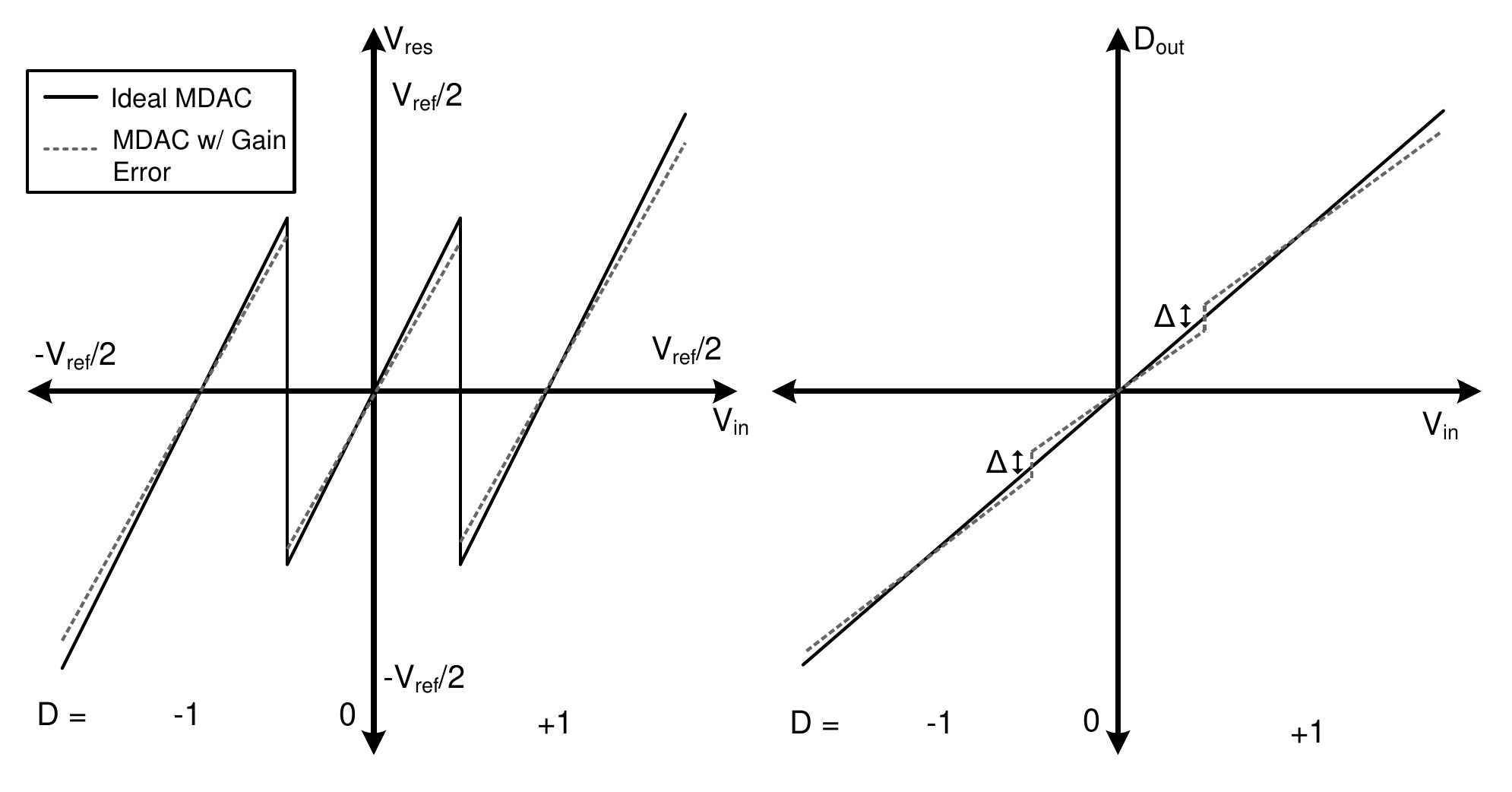}
		\caption{MDAC residue characteristic (left). Reconstructed output (right); $V_{ref}$ - reference voltage; $V_{res}$ - residue voltage; $D$ - subADC decision; $D_{out}$ - reconstructed output.}
		\label{fig:residue_char}
	\end{center}
\end{figure}

Foreground digital calibration is performed to correct for gain errors due to capacitor mismatch~\cite{calib_ref}, exploiting the redundancy offered by the three possible codes produced to resolve one bit. Figure \ref{fig:residue_char} shows the residue output $V_{res}$ of the Stage 1 MDAC and the reconstructed output $D_{out}$, as a function of the input $V_{in}$,  for an ideal MDAC and for an MDAC with gain error. Gain errors due to capacitor mismatch give rise to code jumps in the reconstructed output, as shown in Figure \ref{fig:residue_char}. The calibration algorithm consists of measuring the MDAC code jumps $\Delta$ by subsequent ADC stages and removing them digitally from the reconstructed digital output $D_{out}$. The calibration procedure starts with the last MDAC stage (Stage 4 MDAC in this prototype) and moves backward to calibrate the Stage 1 MDAC. An on-chip control register is used to put the ADC in calibration mode.

\section{Measurement Results} \label{sec:meas_results}

\begin{figure}
	\begin{center}
		\includegraphics[scale=0.5]{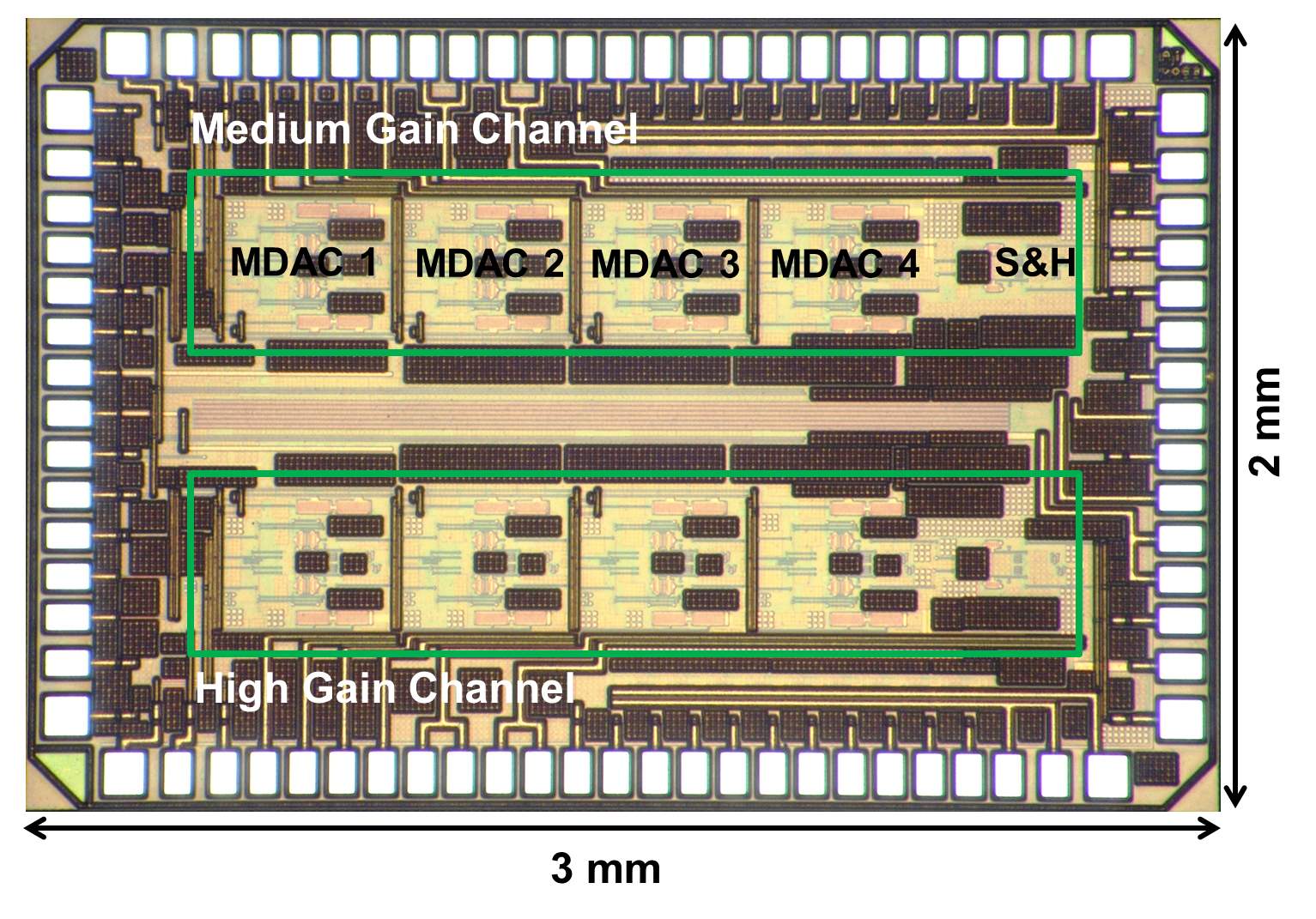}
		\caption{ADC die photograph.}
		\label{fig:die_photo}
	\end{center}
\end{figure}

Figure \ref{fig:die_photo} shows the die photograph of the 2-channel ADC prototype. The chip occupies 6~mm$^{2}$. A 160~MHz sinusoidal clock is fed into the chip and divided down to generate the required 40~MHz clock phases. Going on-chip with a higher frequency clock reduces the jitter contribution due to the clock distribution path. To characterize the dynamic performance of the ADC, an input sinusoidal signal is filtered and fed to the commercial ADC input driver AD8138 which performs single-ended to differential conversion. The signal is split and sent to both the ADC prototype and the commercial ADC (AD9255). The analog residue from the prototype ADC is fed to a different channel of the commercial ADC for further digitization. The digital outputs from the prototype chip and the external ADC are collected by an on-board FPGA and all data processing is done offline.

\subsection{ADC Performance} \label{sec:pre_rad}
\begin{figure}
	\begin{center}
		\includegraphics[scale=0.4]{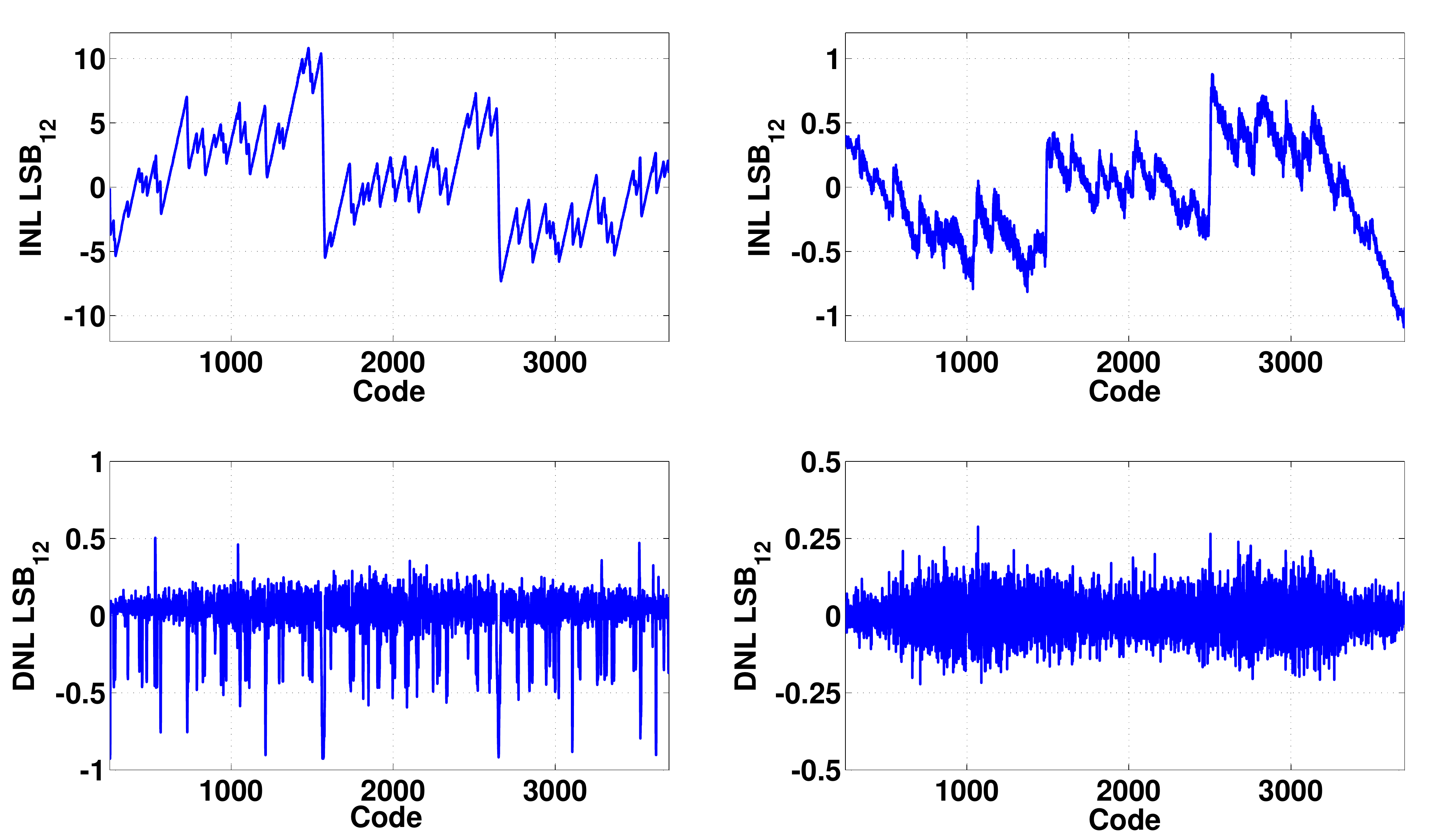}
		\caption{INL/DNL at 40 MS/s ({\it Chip 1}): (left) before calibration, (right) after calibration.}
		\label{fig:inl_full}
	\end{center}
\end{figure}

Both ADC channels were characterized extensively before and after irradiation. 
The static performance of the ADC was measured by the sine-wave histogram method \cite{sine_inl_ref}. Figure \ref{fig:inl_full} shows the static performance of one ADC channel at 40~MS/s before and after foreground digital calibration. The integral nonlinearity (INL) improves from +10/-5 LSB$_{12}$ before calibration to +0.8/-1.1 LSB$_{12}$ after calibration, where LSB$_{12}$ refers to a 12-bit LSB. Similarly, the differential nonlinearity (DNL) improves from +0.5/-0.75LSB$_{12}$ to +0.3/-0.23LSB$_{12}$ after calibration. 

\begin{figure}
	\begin{center}
		\includegraphics[scale=0.4]{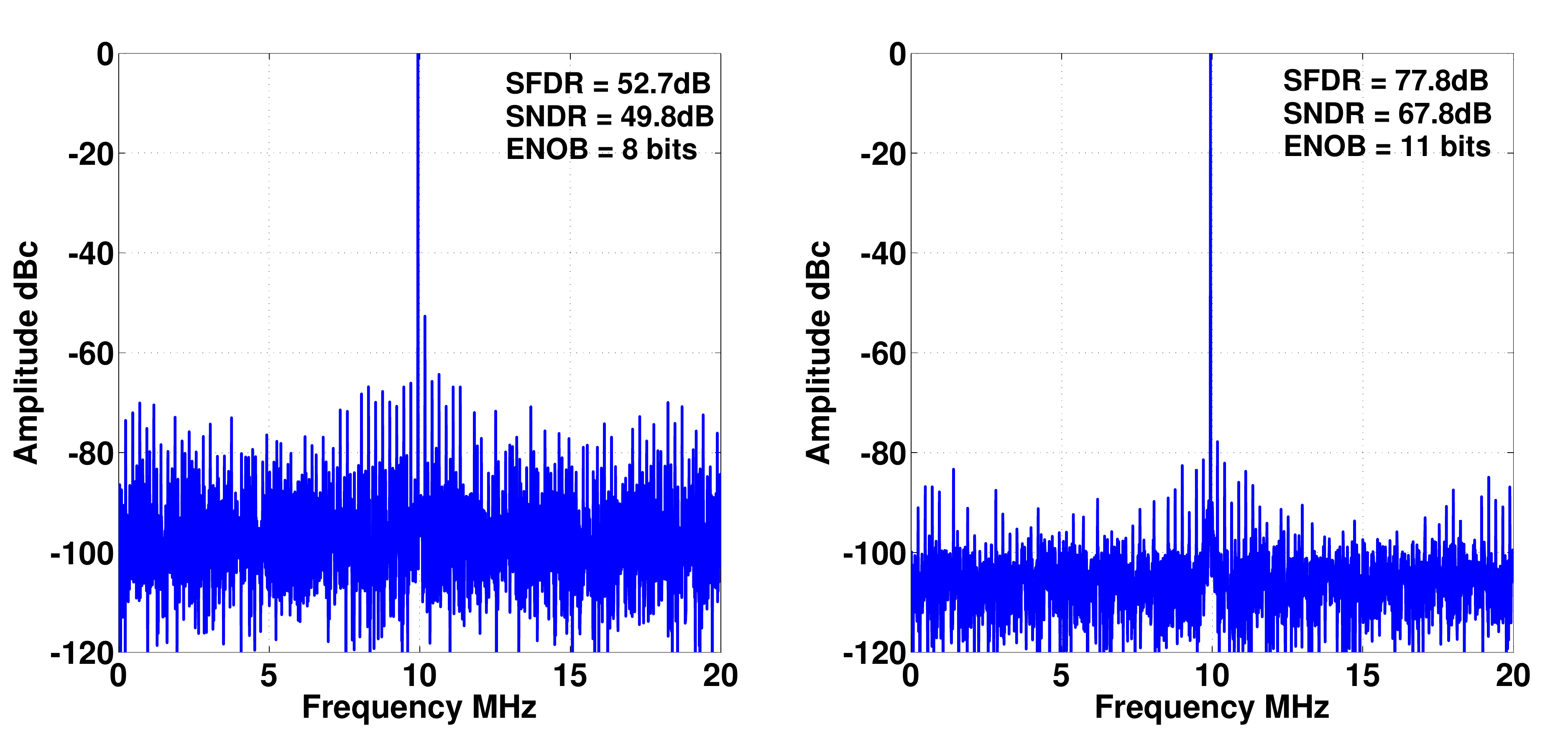}
		\caption{FFT for $f_{in}$ = 10MHz ({\it Chip 1}): (left) before calibration, (right) after calibration.}
		\label{fig:fft_full}
	\end{center}
\end{figure}

\begin{figure}
	\begin{center}
		\includegraphics[scale=0.45]{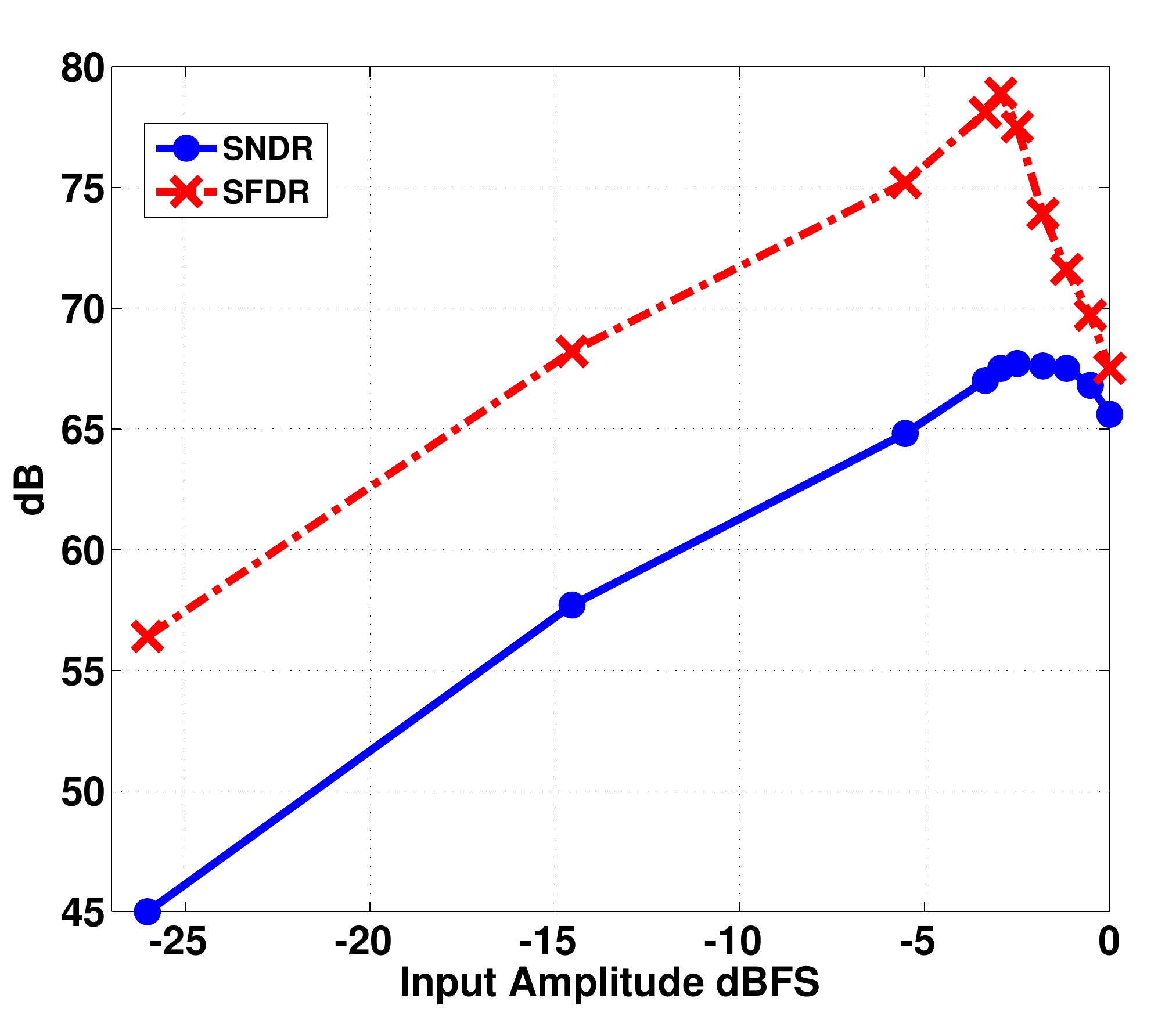}
		\caption{Dynamic performance vs. input amplitude (10MHz) ({\it Chip 1}).}
		\label{fig:ampl_sweep}
	\end{center}
\end{figure}

Figure \ref{fig:fft_full} shows the output Fast Fourier transform (FFT) of a 10~MHz input sine-wave before and after digital calibration. Calibration improves the spurious-free dynamic range (SFDR) from 52.7~dB to 77.8~dB while the signal-to-noise and distortion ratio (SNDR) improves from 49.8~dB to 67.8~dB. The effective number of bits (ENOB) at 10~MHz for the calibrated channel is 11-bits. The number of valid codes was limited to $3452$ as the commercial ADC that further digitizes the analog residue limits the maximum input signal to 85\% of the ADC prototype's full-scale, giving an effective resolution of 11.7-bits. Each MDAC stage consumes 11~mW from a 2.5~V supply. Since all stages are sized identically, the total analog power consumption of the chip is 55~mW per channel (4 identical MDAC stages followed by a Sample and Hold).

Figure \ref{fig:ampl_sweep} shows the SNDR and SFDR as a function of the input amplitude. As mentioned earlier, the valid input range of the on-board commercial ADC limits the maximum allowable input swing to the chip and hence the SNDR peaks at -2~dBFS.

\begin{figure}
	\begin{center}
		\includegraphics[scale=0.35]{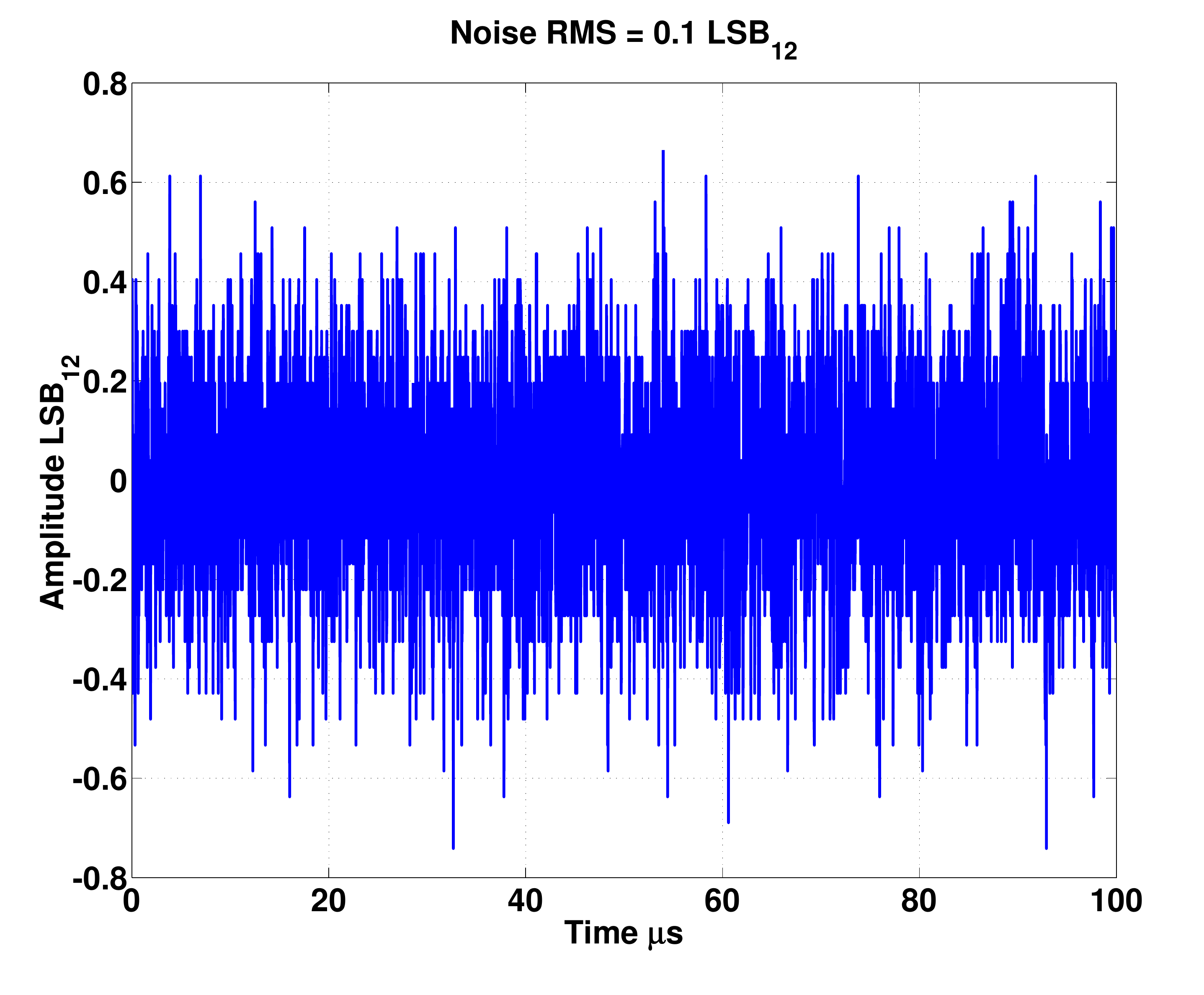}
		\caption{Crosstalk on Medium gain channel.}
		\label{fig:crosstalk}
	\end{center}
\end{figure}

To determine the crosstalk between the two ADC channels, the input to one of the channels was grounded while a full-scale sine-wave was applied to the other channel. Figure \ref{fig:crosstalk} shows the output of the grounded channel. The RMS noise is $0.1$ LSB RMS, showing no indication of crosstalk between the two channels.

\subsection{Irradiation}

\begin{figure}
	\begin{center}
		\includegraphics[scale=1.0]{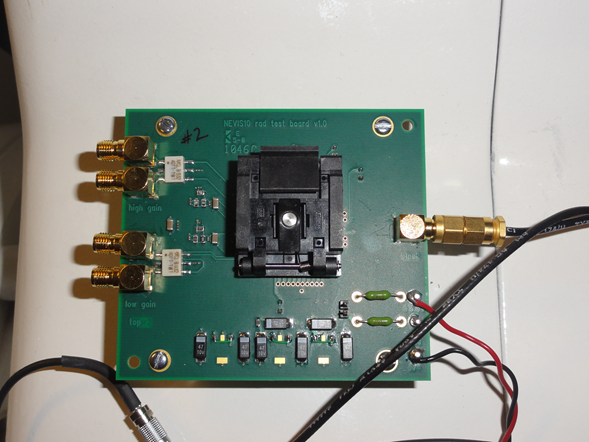}
		\caption{Test board for irradiation.}
		\label{fig:socket_board}
	\end{center}
\end{figure}

\begin{figure}
	\begin{center}
		\includegraphics[scale=0.50]{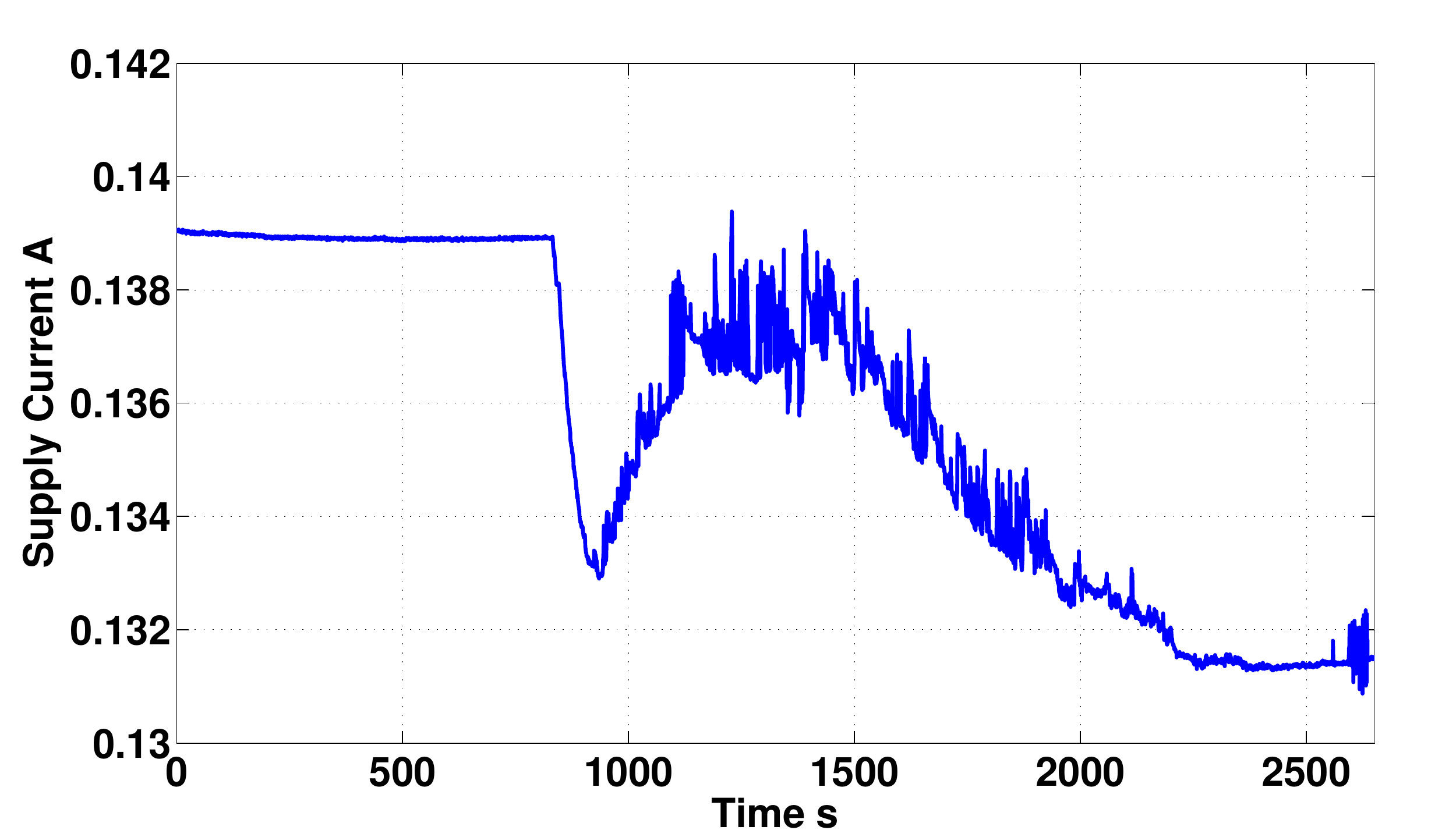}
		\caption{Current consumption variation during irradiation. Note the vertical scale. 2500 s corresponds to a dose of 5 MRad.}
		\label{fig:irr_current}
	\end{center}
\end{figure}

To test the radiation tolerance of this ADC design, five chips were taken to the Massachusetts
General Hospital Francis H. Burr Proton Therapy Center. Two of these (chips 1 and 2) had been fully characterized
and showed excellent performance. The other three were selected using a socketed board (Figure \ref{fig:socket_board}).
This board provided power and a clock signal to the ADC. To confirm that the chips operated as expected, a pure sine
wave signal was given as input (2 V$_{p-p}$) and the analog residue was observed on an oscilloscope.

To replicate the high radiation flux in the ATLAS detector, four of the chips were irradiated using the 227 MeV proton beam. During the irradiation, the chips were powered and a clock signal was applied, while the current drawn by the chip was monitored. Table \ref{tab:irr} lists the dose for the two chips which received the largest dose. No further measurements were done using the three other chips which received smaller doses, given the performance of the high dose chips. In Figure \ref{fig:irr_current}, the current over the course of one test is seen to be relatively constant, between 0.13 and 0.14 A. The drop observed at about 900 seconds shows the beam turn on. The slight rise and fall of the current which follows is due to the heating of the chip as it is irradiated.

\begin{table}
  \begin{center}
    \caption{Measurements of ADC performance before and after irradiation in a 227 MeV proton beam at $f_{in}$~=~10 MHz.}
    \begin{tabular}{c c c c c c}
      \hline \hline
      Chip Number & $\mathrm{proton}/\mathrm{cm}^{2}$ & MRad & SNDR [dBc]  & SFDR [dBc] & ENOB \\
             & &  &  Pre/Post-Irradiation &  Pre/Post-Irradiation &   Pre/Post-Irradiation\\
	\hline	
	1 & $1.01 \times 10^{14} $ & 5.33 & 67.85 / 67.78 & 73.66 / 77.8 & 10.98 / 10.97\\
	2 & $2.01 \times 10^{14} $ & 10.70 & 67.54 / 67.69 & 73.30 / 72.98 & 10.93 / 10.95\\
      \hline \hline
    \end{tabular}
  \label{tab:irr}
  \end{center}
\end{table}

In order to confirm the operation of the ADC immediately after irradiation, a pure sine wave signal was again applied to the ADC input and the analog residue was observed on an oscilloscope. None of the chips ceased operation during the tests. After two months to ensure the chips became safe to handle, the two chips (chips 1 and 2) were re-mounted on the test boards and fully retested.

\subsection{ADC Performance Post-irradiation}
\begin{figure}
	\begin{center}
		\includegraphics[scale=0.45]{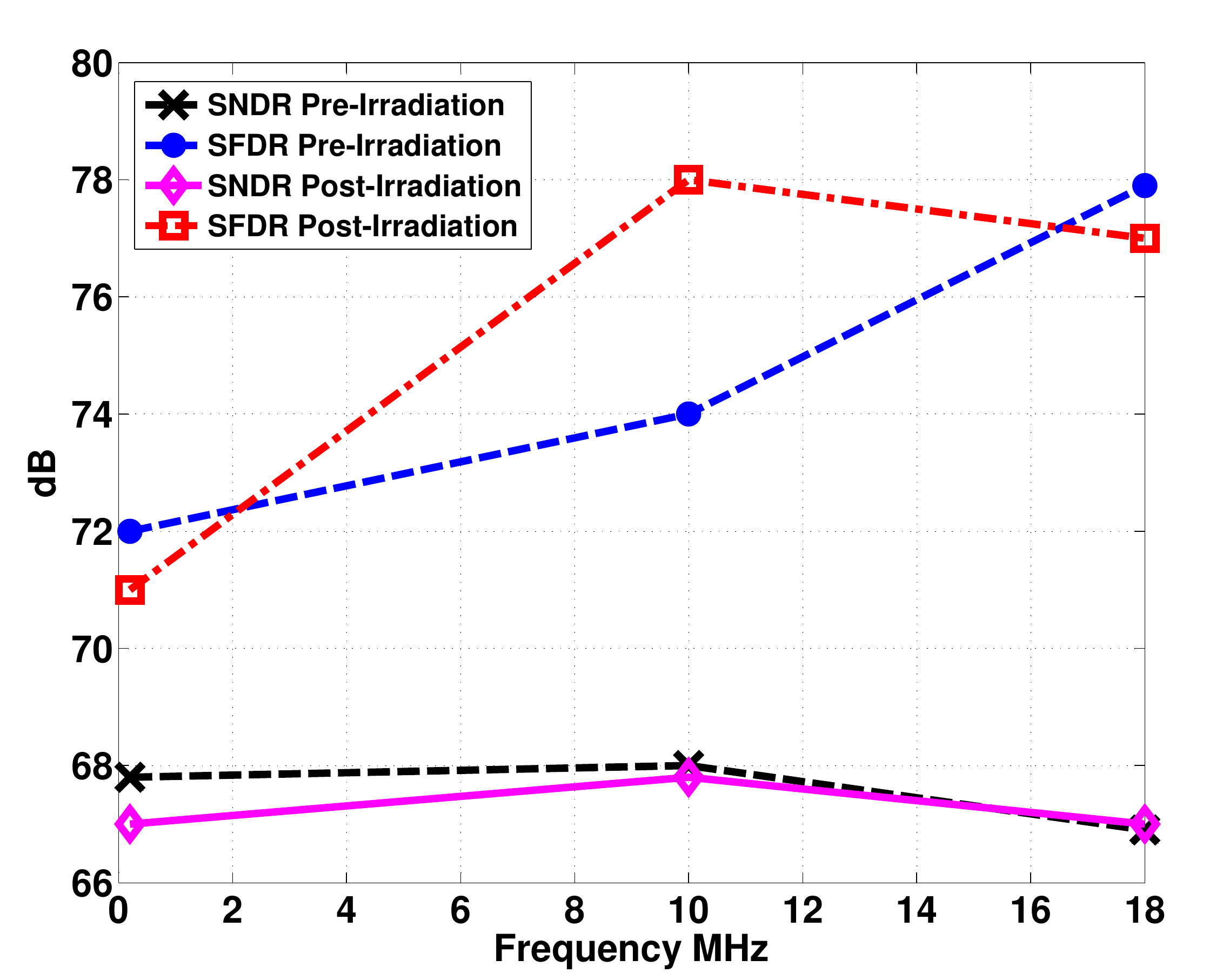}
		\caption{Dynamic performance before and after irradiation (5 MRad) ({\it Chip 1}).}
		\label{fig:comp_irrad}
	\end{center}
\end{figure}

After irradiation the performance measurements were repeated and the results are shown in Table~\ref{tab:irr}. The calibration constants, computed through the digital calibration routine, did not change after irradiation. Figure \ref{fig:comp_irrad} compares the dynamic ADC performance before and after irradiation as a function of the input signal frequency, showing the radiation-hard nature of the design.

\section{Gain Selection} \label{sec:gain_selection}
In the current detector, the 16-bit dynamic range with 12-bit precision requirement is met by using three analog gain channels. The three gain channels (each separated by a factor of approximately ten) are sampled by a $144$-sample deep analog pipeline at 40 MHz~\cite{feb}. An external trigger system identifies the bunch crossing of interest. When such a bunch crossing is identified, five samples for each gain are digitized with 12-bit precision, followed by digital gain selection. A particular channel is said to be saturated if its signal exceeds a certain threshold. Once saturation is detected, any signal from that particular channel is ignored for $\sim$~500~ns (due to the recovery time required for the conditioning circuits in the signal path) and the data from the next lower unsaturated gain channel is used.

For the upgrade, there will be no external trigger system and the data for all samples must be transmitted. This requires the ADC to sample the input at 40 MS/s and the optimal gain to be chosen on the fly.

A simple solution uses a single discriminator (comparator) on the sampled signal in front of the highest gain channel, which imposes a switch to a lower gain input if the signal is too close to saturation (based on a single sample). This approach requires two sample and holds at the ADC input, one for a $1\times$, medium gain channel and one for a $10\times$, high gain channel, resulting in a sample depth of one (the OTA for the Stage 1 MDAC can be shared between the two sample and holds). This implementation of gain selection leads to limited precision at the onset of large signals, due to the limited bandwidth of the sampling network and the shape (width) of the signal. The signal from a saturated (high gain) channel can have a very high slew rate, which may result in a large error for the sampled value. Since the value for the sampling capacitor is set by $kT/C$ noise requirements \cite{analog_ref}, the size of the sampling switches determines the input bandwidth. The switches in the ADC input sampling network are sized to sample a full-scale sine-wave signal at the Nyquist frequency of 20 MHz. Larger switches increase power consumption on the clock drivers. The maximum slew rate of the input signal that can be sampled with the required 12-bit accuracy is given by

\begin{equation}
        \mbox{Maximum Slew Rate } SR_{max} = V_{FS}(2\pi f_{max})
\end{equation}
where $V_{FS} = 1.2$~V, $f_{max} = 20$~MHz for the current design, and therefore $SR_{max} = 1.5 \times 10^8$~V/s.

\begin{figure}
	\begin{center}
		\includegraphics[scale=0.4]{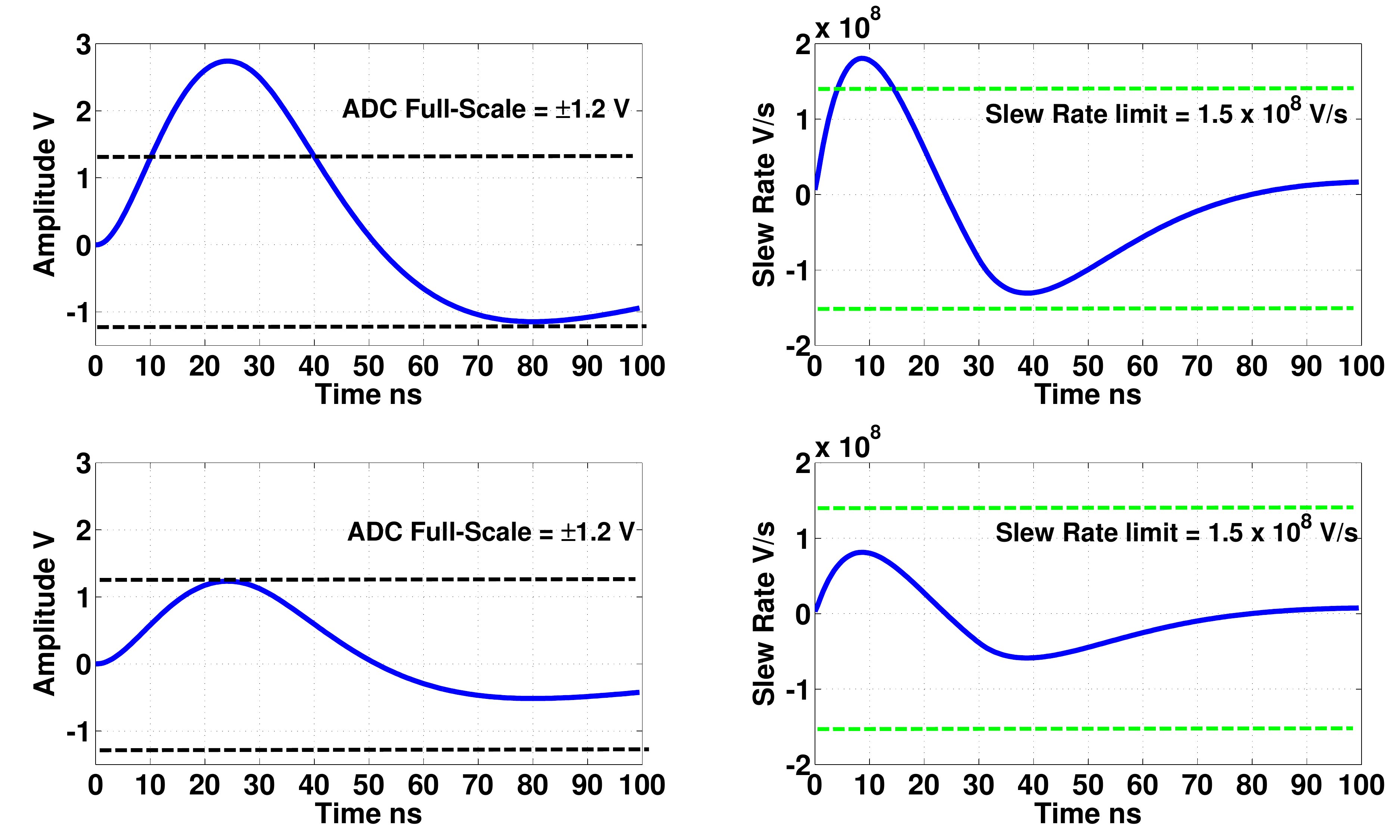}
		\caption{High gain channel output with a highly saturated pulse (top-left), and corresponding slew rate (top-right). A slightly saturated pulse (bottom-left) and its slew rate (bottom-right) are shown for comparison. }
		\label{fig:sat_pulses}
	\end{center}
\end{figure}

Figure \ref{fig:sat_pulses} shows two different pulse shapes with their respective slew rates. For simplicity, the input signal is assumed to be saturated if it is larger than the ADC full-scale of $\pm$1.2~V. The signal is said to be slightly saturated if it just goes above the ADC full-scale, and highly saturated if it is very much larger. In practice, protection diodes prevent the ADC input signal from going much beyond the supply voltage of 2.5~V. The value of the highly saturated signal at 10~ns in Figure~\ref{fig:sat_pulses} is still {\it within} the ADC full-scale range. However, the slew rate of the highly saturated pulse is larger than the maximal slew rate supported by the input sampling network $SR_{max}$. Therefore, the sampled value will be inaccurate. This illustrates how the solution of using a single comparator, at the input sample and hold, to detect saturation and then switching between channels, leads to limited precision. 

A second solution is to switch to the lower gain if the signal saturates or the slew rate is too large. To implement this solution the comparator described in the previous case can be used, but circuitry to determine the slew rate of the input signal needs to be added. If, at the sampling instant, the slew rate of the input signal is higher than a threshold, the lower gain channel samples are used for digitization. Since the slew rate is only used to switch between the gains, it does not need to be measured with 12-bit accuracy. However, it does determine the point where the gains switch, and thus the usable part of the dynamic range of the high gain channel. To ensure the correct gain channel is used, the switch to a lower gain should occur well before the signal in the high gain channel comes close to its maximum.  The point at which this switch happens can be determined as a function of the slew rate accuracy and threshold. With a lower slew rate accuracy, a lower threshold is needed and hence more of the dynamic range of the higher gain channel is lost. These parameters could be adjusted for optimal performance. Due to the inherent pipeline delay, and the fact that adjacent input samples are 25~ns apart (the sampling clock is 40~MHz), a digital approach to determine the input signal slew rate is not feasible. Determining the slew rate in an analog fashion (using differentiators) with the necessary accuracy would require complex circuitry and could significantly increase the power consumption.

A simple and robust solution, which is independent of the signal shape and slew rate, is to detect saturation with a comparator in front of the first stage of the high gain channel (as before).  When saturation is detected, the $N$ samples in high gain channel stages 1 to $N$ are ignored. The lower gain channel is then used to digitize the $N$ samples before saturation was detected, as well as the future samples. This procedure requires a memory depth of $N > 1$ (as shown in Figure~\ref{fig:GS_block}). By ignoring the $N$ samples before saturation in the high gain channel, the samples that may suffer from limited precision due to the high slew rate are avoided. 

\subsection{Gain Selection Measurements} \label{sec:gs_results}
In the prototype chip, two discriminators are implemented for each gain (00: signal saturated with negative value, 01: signal within range, 11: signal saturated with positive value). The discriminator memory is implemented outside the chip for flexibility in the study of the gain selection algorithm.  It is necessary to determine the minimum depth of the discriminator memory required for the gain selection algorithm. As the gain selection is made only after a certain number of sampling clock cycles, the length of discriminator memory defines the length of the parallel analog pipelines for the signal path, i.e.~the number of parallel MDACs which must exist in the final chip to do the gain selection. 

\begin{figure}
	\begin{center}
		\includegraphics[scale=0.5]{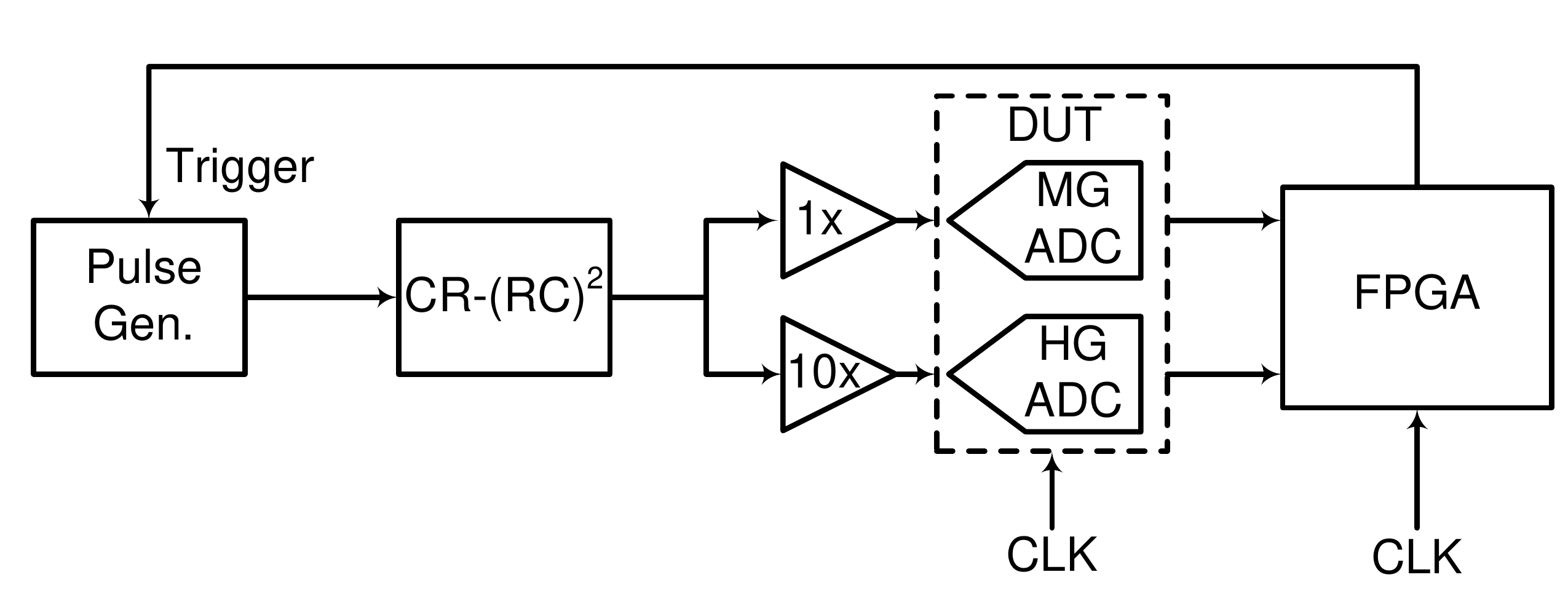}
		\caption{Measurement setup for gain selection.}
		\label{fig:gs_setup}
	\end{center}
\end{figure}

The setup used for the gain selection experiments is shown schematically in Figure \ref{fig:gs_setup}. A pulse generator is used to generate an approximation of the signal from a calorimeter cell. The pulse is then subjected to a fast bipolar $CR-(RC)^{2}$ shaping and injected into the medium and high gain channels through two amplifiers. As shown in Figure \ref{fig:chip_arch}, each ADC channel has two comparators at its input to perform gain selection. 

\begin{figure}
	\begin{center}
		\includegraphics[scale=0.5]{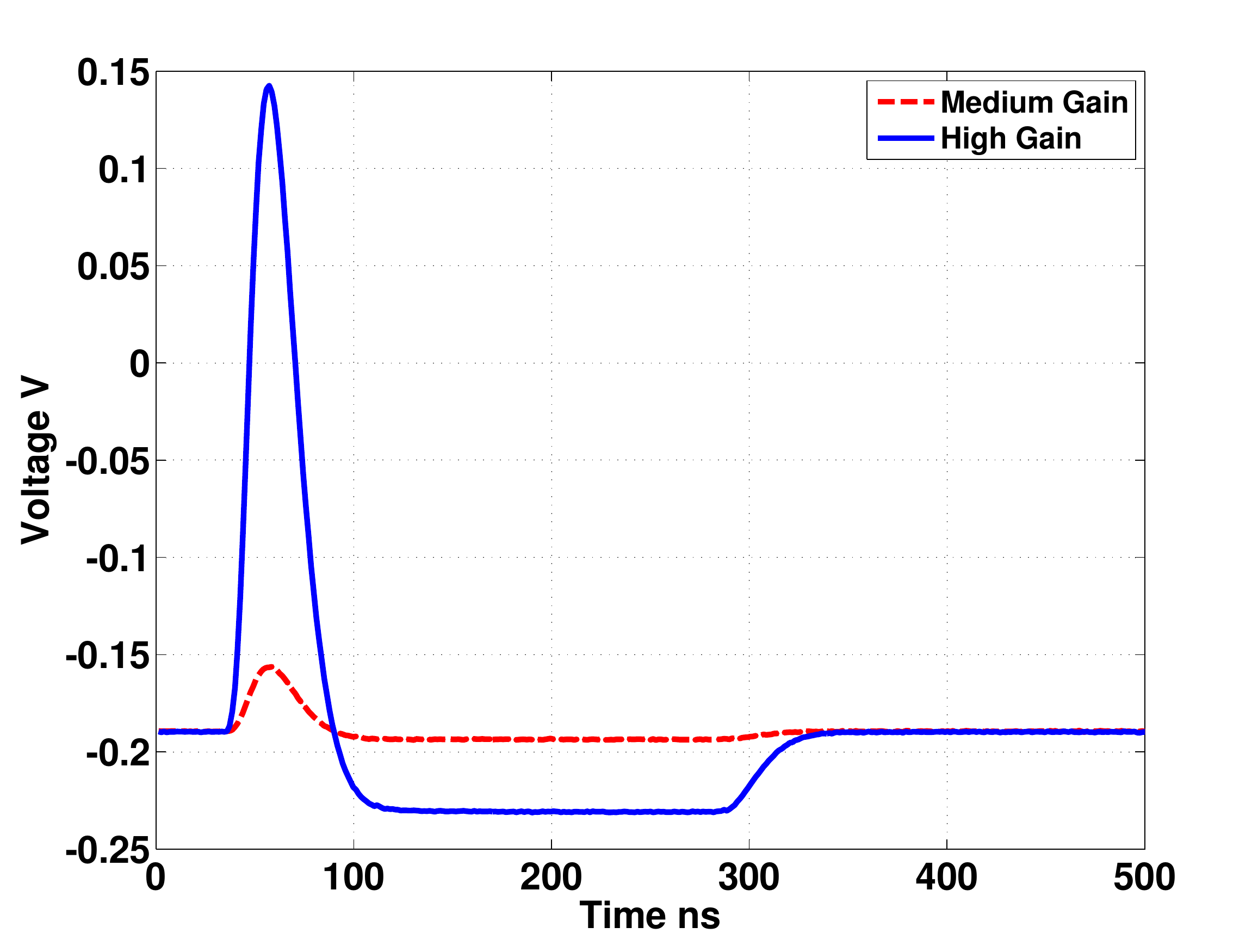}
		\caption{Reconstructed medium and high gain pulses.}
		\label{fig:meas_pulse}
	\end{center}
\end{figure}

For the gain selection study, the phase of the pulse generator was synchronized to the ADC sampling clock through the FPGA, with an option to change the phase of the input signal with respect to the sampling clock, making a fine reconstruction of the pulse shape possible. Figure \ref{fig:meas_pulse} shows reconstructed pulse shapes measured by the medium-gain and high gain channels with $\tau$~=~RC~=~20~ns.

To verify the operation of the gain selection setup and to confirm the depth of the discriminator memory required to perform the gain selection detailed above, the following procedure was followed:

\begin{itemize}
	\item The output of the high gain channel is collected continuously and the gain selection comparators are used to determine if the high gain is in saturation.
	\item Without the signal, the background mean $\mu$ and standard deviation $\sigma$ of the channel output (in terms of the ADC LSBs) are determined. The start of the pulse is defined as the first sample which is $\mu + n\sigma$ above the background, where $n$ is varied from one to three.
	\item If the high gain channel is saturated, the number of samples from the start of the pulse until saturation is noted. For the simpler approach, this gives the depth $N$ of the required discriminator memory and hence the depth of the required analog pipeline (number of parallel MDACs).
\end{itemize}

\begin{figure}
	\begin{center}
		\includegraphics[scale=0.42]{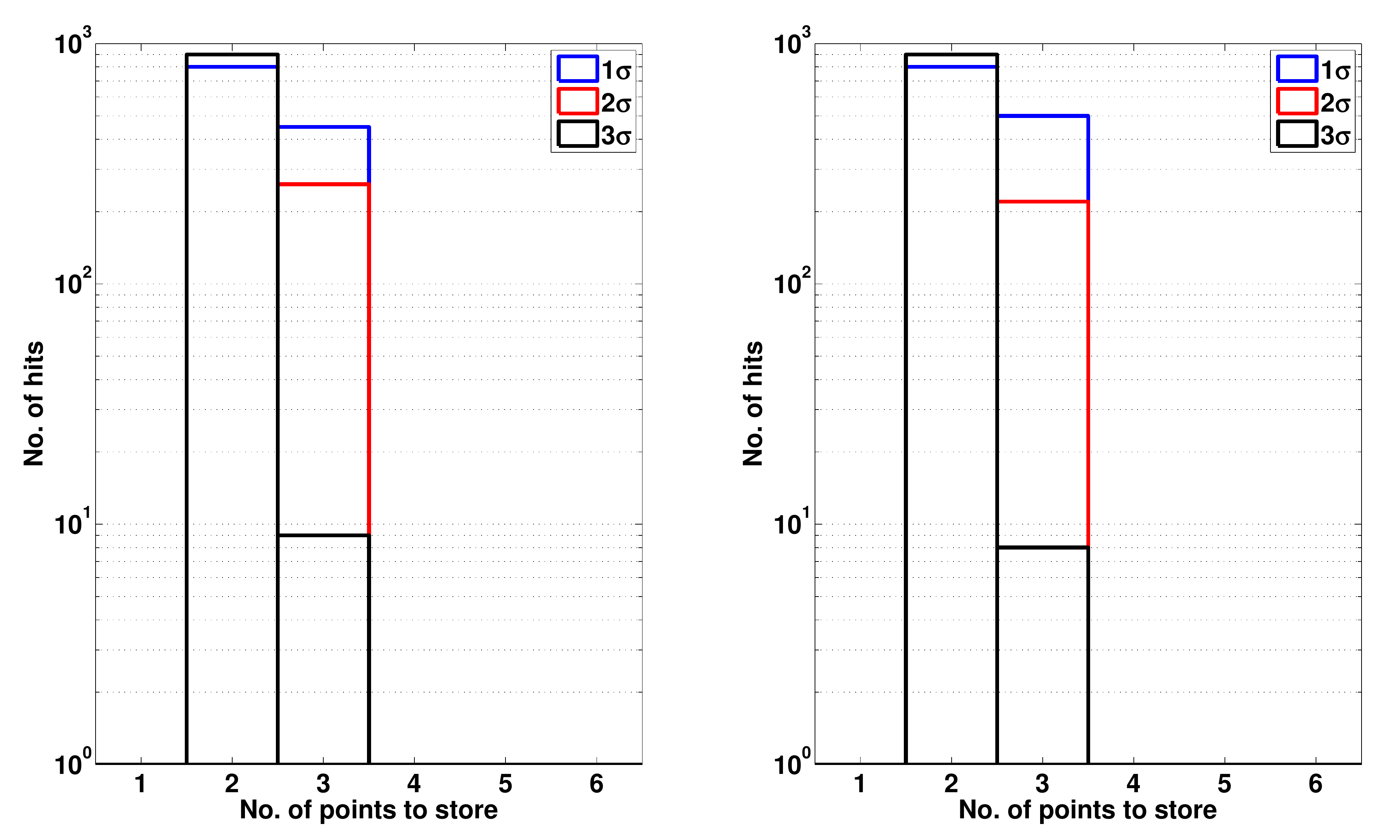}
		\caption{Required memory depth for a highly saturated signal when pulse and sampling clock are: (left) in phase, (right) out of phase by 12.5~ns. The curves are labeled by $n \times \sigma$, the threshold used to determine the start of the pulse.}
		\label{fig:highly_sat}
	\end{center}
\end{figure}

In the first experiment, a high amplitude signal with a fast rise time was applied to the high gain channel to saturate it fully. This is similar to the situation in the top row of Figure~\ref{fig:sat_pulses}. Figure~\ref{fig:highly_sat} shows the required memory depth for two different phase relationships between the pulse shape and the ADC sampling clock. As can be seen from the figure, at most three points need to be stored.

\begin{figure}
	\begin{center}
		\includegraphics[scale=0.42]{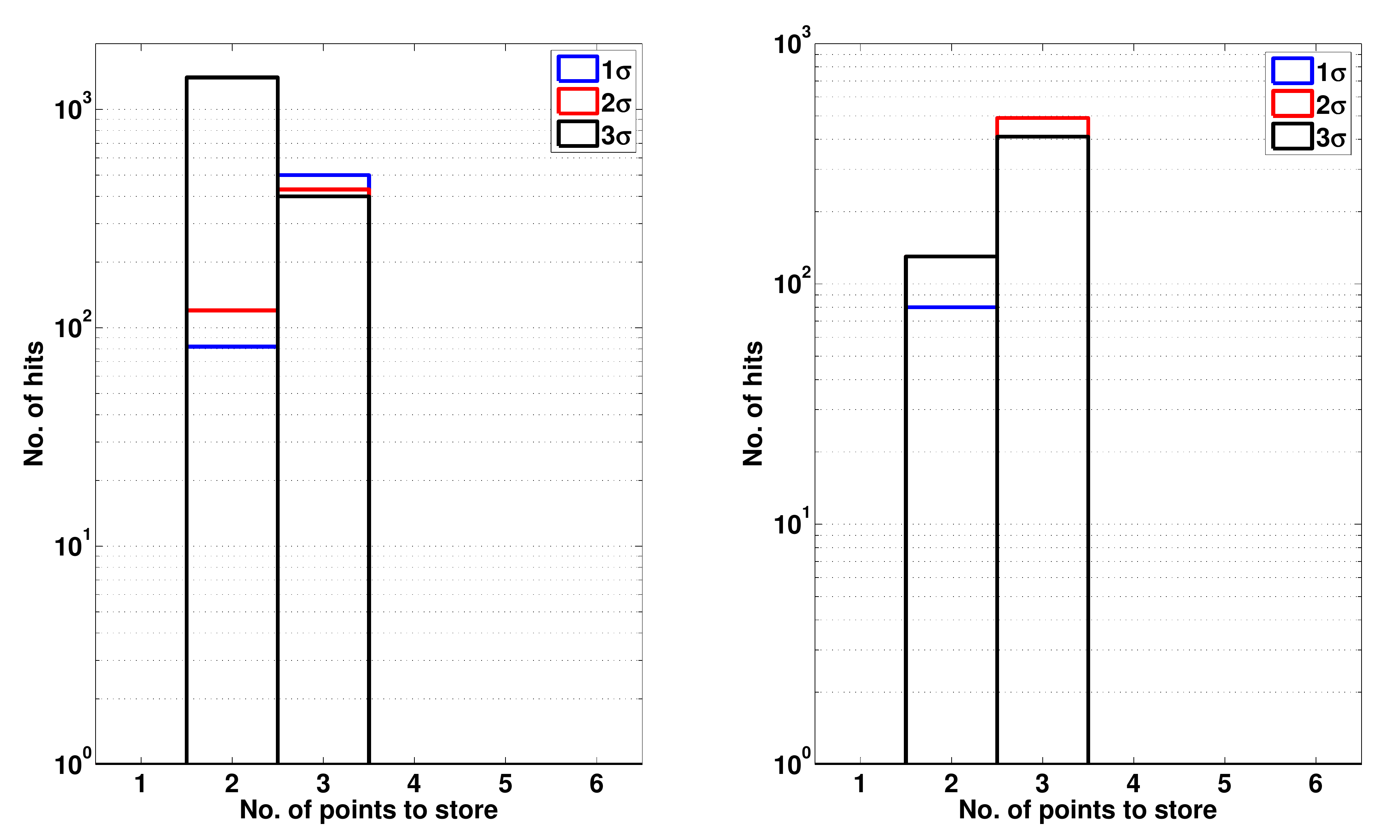}
		\caption{Required memory depth for a slightly saturated signal when pulse and sampling clock are: (left) in phase, (right) out of phase by 12.5~ns.}
		\label{fig:slightly_sat}
	\end{center}
\end{figure}

In the next experiment, a signal was applied to the high gain channel with an amplitude that was just over the saturation point of the high gain channel (as in the bottom row of Figure~\ref{fig:sat_pulses}). Figure \ref{fig:slightly_sat} shows the required memory depth for two different phase relationships between the pulse shape and the ADC sampling clock. As before, the required memory depth is three samples. From Figures \ref{fig:highly_sat} and \ref{fig:slightly_sat}, the maximum memory depth required for $\tau$~=~RC~=~20~ns is three samples.

\begin{figure}
	\begin{center}
		\includegraphics[scale=0.42]{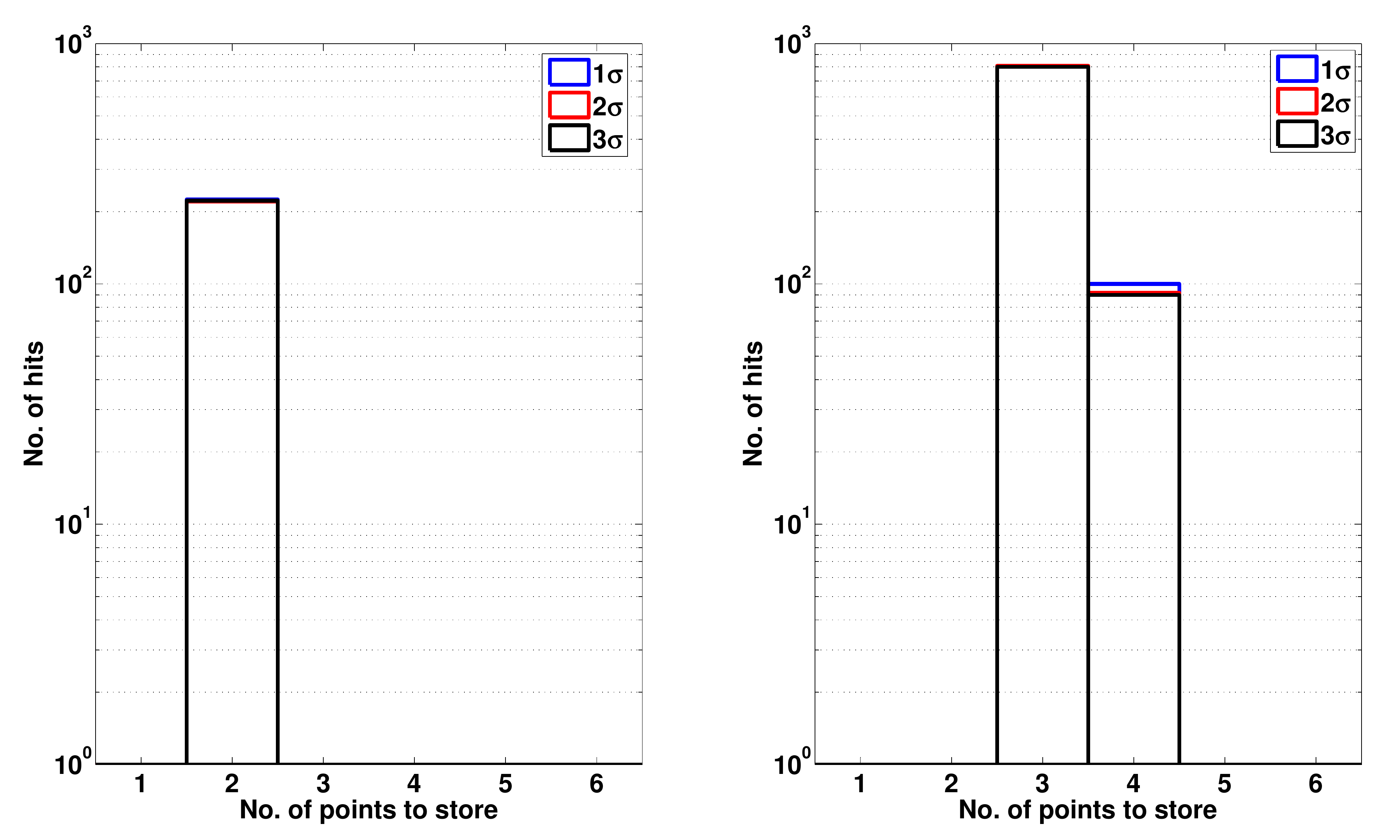}
		\caption{Required memory depth for: (left) $\tau = 10$~ns: (right) $\tau = 40$~ns.}
		\label{fig:diff_tau}
	\end{center}
\end{figure}

The time constant $\tau$ of the pulse can vary from one calorimeter cell to another due to variations in the cell capacitances. Figure \ref{fig:diff_tau} shows the required depth for two different pulse shapes with $\tau = 10$~ns and $\tau = 40$~ns. The maximum required memory depth can be as high as four samples. These tests confirm that a memory depth of four samples is sufficient to avoid problems associated with slew rates of the input signal and to perform accurate gain selection.

\subsection{Gain Selection Algorithm}
Several possibilities for performing the gain selection are:

\begin{enumerate}
	\item {\bf Fully Digital Gain Selection}:  Implement three complete ADC channels, one for each of the three gains, with the gain selection performed after the digitization is complete. The final ADC  word that is transmitted contains the digital word from the appropriate channel and its gain scale.  This leads to three independent 12-bit ADC channels per 16-bit dynamic range. For a quad-ADC, this adds challenges in terms of crosstalk, power and clock distributions.

	\item {\bf Partial Digital Gain Selection}: Implement three partial ADC channels, sharing the back-end stages using an analog multiplexer. In this case, each gain channel would have a certain number of independent MDACs stages, followed by an analog multiplexer that enables the three channels to share the back-end stages. In order to meet the 16-bit dynamic range requirement, three partial channels, each with four independent MDACs followed by an analog MUX and a common back-end, are required to perform the gain selection.

	\item {\bf Single 16-bit Dynamic Range ADC}: Implement one 16-bit dynamic range ADC channel with 12-bit precision. The ADC noise performance needs to be 16-bit accurate while the linearity needs to be only 12-bit accurate. This solution avoids gain selection entirely as there is only a single ADC channel. To meet the 16-bit noise requirements, the input sampling capacitance of the ADC would need to be $> 70$~pF, presenting a very large capacitive load to the ADC input signal driver. Also, the power of the first stage MDAC would need to be increased dramatically to achieve the required 16-bit performance.  However, the additional power required by the 16-bit channel over a 12-bit channel is offset by reducing the number of external amplifiers and full or partial ADC channels from three to one.
\end{enumerate}

The primary advantage of the fully digital approach is its flexibility.
In the partial channels approach, the memory depth depends on the signal shape. Since in a conventional pipeline ADC successive pipeline stages are progressively smaller and lower power, the amount of area and power savings provided by reusing the back-end stages in the partial channels approach is small. Furthermore, the chip complexity is not very much reduced when compared to the fully digital approach.

\section{Conclusion} \label{sec:Conclusion}
The design of a radiation-hard dual-channel 12-bit 40 MS/s Pipeline ADC with extended dynamic range was presented, for use in the readout electronics upgrade for the ATLAS Liquid Argon Calorimeters at the CERN LHC. The ADC was confirmed to be radiation tolerant beyond the required specifications. Various gain selection experiments were performed with the prototype to investigate possible gain selection procedures for future implementations of this design.

\acknowledgments
We would like to acknowledge the excellent work and essential contributions of the technical staff at Nevis Laboratory. This work has been supported by the US National Science Foundation, award number 1067934 (Columbia).

\end{document}